\documentclass[aps,twocolumn,superscriptaddress]{revtex4}
\usepackage[utf8]{inputenc}
\usepackage{graphicx,color}
\usepackage{dcolumn}
\usepackage{bm}
\usepackage{color}
\usepackage{amsmath}
\usepackage{braket}
\usepackage{comment}
\usepackage{bbold}
\usepackage{amssymb}
\usepackage{enumerate}
\usepackage{amsthm}
\usepackage{commath}
\usepackage{natbib}
\DeclareMathOperator{\Tr}{Tr}

\newtheorem{theorem}{Proposition}

\begin{document}


\title{Self--averaging of random quantum dynamics}

\author{Marcin Łobejko}
\affiliation{Institute of Physics, University of Silesia in Katowice,  Katowice, Poland}
\affiliation{Silesian Center for Education and Interdisciplinary Research, University of Silesia in Katowice,  Chorz{\'o}w, Poland}
 \author{Jerzy Dajka}%
\affiliation{Institute of Physics, University of Silesia in Katowice,  Katowice, Poland}
\affiliation{Silesian Center for Education and Interdisciplinary Research, University of Silesia in Katowice,  Chorz{\'o}w, Poland}
\author{Jerzy Łuczka}
\email{jerzy.luczka@us.edu.pl}
\affiliation{Institute of Physics, University of Silesia in Katowice,  Katowice, Poland}
\affiliation{Silesian Center for Education and Interdisciplinary Research, University of Silesia in Katowice,  Chorz{\'o}w, Poland}


\begin{abstract}

Stochastic dynamics of a quantum system  driven by $N$ statistically independent  random sudden quenches in a fixed time interval  is studied. We reveal that  with growing $N$  the system approaches a deterministic limit indicating self-averaging with respect to its temporal unitary evolution. This phenomenon is quantified by the variance of the unitary matrix governing the time evolution of a finite dimensional quantum system which according to an asymptotic analysis decreases at least 
 as $1/N$. 
For a special class of protocols (when the averaged Hamiltonian commutes at different times), we  prove  that for  finite  $N$  the distance (according to the Frobenius norm) between the averaged unitary evolution operator generated by the Hamiltonian $H$ and  the unitary evolution operator generated by the  averaged  Hamiltonian  $\langle H \rangle$ scales as $1/N$. Numerical simulations enlarge this result to a broader class of the non-commuting protocols.    
\end{abstract}


%
\maketitle


\section{Introduction}
\emph{Self--averaging}  is a well established concept in  statistical physics of disordered and random systems. 
Loosely speaking, a certain property $X$ 
of a system is self-averaging if most realizations of the randomness have the same value of $X$  in some limiting regime. More precisely, a system is self-averaging
with respect to  $X$  if the relative variance of $X$ tends to zero in this limiting regime. If e.g. we consider a system of combinatorial objects of size $N$
then the relative variance 
\begin{equation}  \label{1} 
 \frac{\langle X_N^2 \rangle - \langle X_N \rangle^2}{\langle X_N \rangle ^2} 
 \to 0
 \end{equation}
 as $N\to\infty$. 
For a large class of randomly driven quantum systems such as \emph{quenched disordered systems}~\cite{quenched, quenched1}, the question about self--averaging of their properties is essentially non--trivial ~\cite{lattice}.  There have been  studies on  self--averaging of a free energy for spin systems with short--range~\cite{short} or long--range interactions~\cite{long}, self--averaging of diffusion in heterogeneous media ~\cite{subdiffusion},  self-averaging of Lyapunov exponents in fluids \cite{Das} and self-averaging of the reduced density matrices \cite{PRA17} to mention only a few. 

In the paper,  we consider a broad class of randomly driven quantum systems for which the  time evolution  is universally self-averaging. In particular, we study  quantum dynamics in the presence of a sequence random and independent step-like perturbations of finite-dimensional quantum systems.  
Such a driving corresponds to  \emph{quantum quench} dynamics of closed quantum systems --  a rapidly developing and intensively investigated research area~\cite{essler}  which recently has found  experimental realizations \cite{exper}. 
 Thermalization \cite{thermal}, quantum phase transitions \cite{QPT,QPT1}, integrability \cite{integra} and simple out-of-equilibrium quantum systems \cite{noneq} - it is a far from complete list of examples where  quantum quench scenarios have 
  been  studied.  
We investigate the driving of a quantum system formed as a series of statistically independent random quenches -- \emph{multiple random quench} (MRQ) and its  continuous limit of an infinite number of quenches occurring in a finite time interval --  \emph{continuous random quench} (CRQ). Self-averaging of the unitary  time evolution for the MRQ protocol occurs with increasing number $N$ of quenches in the fixed time interval. This phenomenon is quantified by vanishing variance of the unitary time-evolution matrix representation that decreases at least as $1/N$.  This behaviour is  formally proved for an arbitrary distribution supported on bounded intervals of the randomly controlled Hamiltonians. 
According to the self-averaging property,  the considered unitary evolution converges almost surely to its mean value. We estimate this mean value for a special class of protocols when an instantaneous average of the Hamiltonian (with respect to the matrix ensemble) commutes at different time instants. We call this property 'the commutation in the statistical sense'. For this case we prove that the self-averaged unitary evolution converges  to the evolution governed by a mean value of a random Hamiltonian and convergence is in the sense of the Frobenius (Hilbert-Schmidt) norm. In other words, in the basis where the average of the Hamiltonian is diagonal, off-diagonal elements with vanishing mean value less and less contribute to the time evolution as a number of quenches increases. 
Moreover, we have also performed numerical simulations in order to analyze  a non-commuting case for a qubit. For some particular drivings we show that also in this case, in the CRQ limit, the evolution is generated by a mean value of the Hamiltonian even though it does not commute in a statistical sense at different time instances (i.e. when instantaneous averages cannot be simultaneously diagonalized). For this non-commuting case and two other examples of the MRQ protocols for a qubit space, results of numerical simulation apparently exhibit the exact power law  $1/N$  which  is the lower asymptotic estimation predicted analytically.

The layout of the paper is as follows. In Sec. II,  we provide a necessary information on theory of  random matrices  required for further reasoning. Next, in Sec. III, we formulate a unitary time evolution of quantum systems with random quenches and introduce the notion of the effective Hamiltonian of the system. In the same section, we define commutation of operators in the statistical sense.    In Sec. IV, we discuss the statistics of the effective Hamiltonian of the MRQ control (with two main propositions concerning its properties) and as a consequence we formulate a self-averaging condition for the unitary time-evolution. In Sec. V, we provide a numerical simulation for more general MRQ protocols. 
Finally, in Sec. VI,  we summarize our results and we present some ideas for future work. We postpone proofs of the propositions formulated in Sec. IV  to  Appendices. 

\section{Random matrix theory}

In order to  describe and define MRQ  we utilize Random Matrix theory \cite{mehta}, a rapidly developing branch of mathematics useful in many branches of modern physics starting from Wigner's classification of ``canonical'' random matrix ensembles for the description of  statistics of nuclear levels spacing up to quantum chaos, many--body physics and quantum statistical mechanics. The MRQ driving studied in this paper is a further example.
 
Let us represent an $M$-dimensional complex and Hermitian matrix $H$ as a point $H = (h_1, h_2, \dots, h_d)$ in a $d$-dimensional real space $ \mathbb{R}^{d}$ where $d = M^2$ is the number of real and independent parameters  specifying the matrix $H$.
In the following, we consider an ensemble of matrices with  {\it random} parameters $h_i$ 
and the probability distribution  
  \begin{eqnarray}\label{mmF}
\text{Pr}(H \in D) = \int_D dH \varrho(H) 
\end{eqnarray}
that   $H = (h_1, h_2, \dots, h_d)\in D \subset \mathbb{R}^d$, 
where $\varrho(H) = \varrho(h_1, h_2, \dots, h_d)$ is a probability density function (pdf) and $dH = dh_1 dh_2 \dots dh_d$. We restrict our reasoning only to the distribution $\varrho(H)$, which we call a matrix-pdf, supported on the bounded probability space $\mathcal{P} \subset \mathbb{R}^d$ and normalized in such a way  that 
\begin{eqnarray}
\int_\mathcal{P} dH \varrho(H) = 1 \,. 
\end{eqnarray}
Let $\mathcal{H}_N$ be an ordered set of random and statistically independent matrices
\begin{equation}
    \mathcal{H}_N = (H_1, H_2, \dots, H_N), \,
\end{equation}
with the joint pdf given by the product of individual distributions ensuring    statistical independence, 
\begin{equation} \label{pdf}
    \rho(\mathcal{H}_N) = \varrho_1(H_1)\varrho_2(H_2)\dots\varrho_N(H_N),   \,
\end{equation}
where the pdf $\varrho_k(H_k) = \varrho_k(h_1^{(k)}, h_2^{(k)}, \dots, h_d^{(k)})$ for $k=1, 2, \dots,N$. 
For any matrix  $U$ depending on the set $\mathcal{H}_N$ one can define the first statistical moment $\braket{U}$ as an average of the elements $[\braket{U}]_{\alpha \beta} = \braket{[U]_{\alpha \beta}}$, where
\begin{eqnarray} \label{avg}
 \braket{[U]_{\alpha \beta}} = \int_\mathcal{P} d\mathcal{H}_N \; [U]_{\alpha \beta} \ \rho(\mathcal{H}_N).  \,
\end{eqnarray}
Here,  $[ \cdot ]_{\alpha \beta}$ denotes a matrix element, $d\mathcal{H}_N = \prod_{k=1}^N dH_k$ and  $dH_k = dh_1^{(k)} dh_2^{(k)} \dots dh_d^{(k)}$. We define {\it per analogiam} a variance-matrix $\mathrm{Var}(U)$ as a matrix of variances i.e. $[\mathrm{Var}(U)]_{\alpha \beta} = \mathrm{Var}([U]_{\alpha \beta})$ with 
\begin{eqnarray} \
 \mathrm{Var}([U]_{\alpha \beta}) = \braket{|[U]_{\alpha \beta}|^2} -|\braket{[U]_{\alpha \beta}}|^2.   \,
\end{eqnarray}
 In the following we use a Frobenius matrix--norm $\| \cdot \|$ defined by
\begin{equation}\label{norm}
\| U \|^2 = \Tr[UU^\dag] \,
\end{equation}
which is known to be  sub-multiplicative,   i.e. $\| A B \| \le \|A\| \|B\|$ for any matrices $A$ and $B$.

\begin{figure*}[t]
\includegraphics[width=\textwidth]{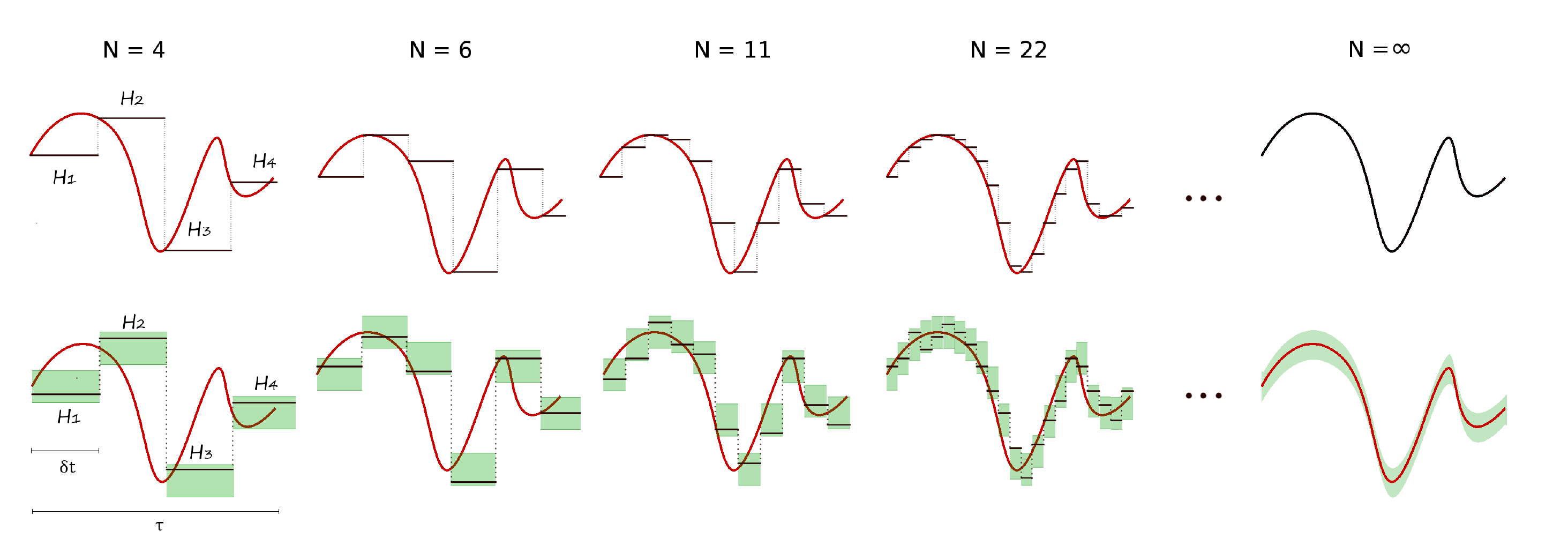}
\caption{Schematic visualization of the  process of ``integration'' of arbitrary time-dependent driving into the unitary time-evolution operator. Solid red lines represent  changes in time of an arbitrary finite-dimensional time-dependent Hamiltonian. 
The deterministic regime in the upper panel shows  an increasing number of partially constant Hamiltonians (solid black lines) forming a step-like realization which in the limit of an infinite number of quenches converges into a time-evolution operator of a continuous driving (red line). The stochastic realization in the lower panel is a sketch of the same idea of an ``integration'' into unitary operators, however, in this case for each partially-constant Hamiltonian we take a statistically independent random matrix with some dispersion (indicated by green boxes). In this case  black solid lines represent a particular realization of this stochastic process and the red line represents a mean value of the time-dependent Hamiltonian.} 
\label{f1}
\end{figure*}

\section{Sudden quench evolution}
In this section,  using the random matrix terminology,  we formulate time--evolution of quantum systems subjected to random quenches. For completeness we start with the more intuitive case of deterministic dynamics which can be considered as a limiting case of more general dynamics which is our primary object of investigation. 

\subsection{Deterministic case}

We consider a  quantum system driven  by a deterministic time-dependent Hamiltonian $H(t)$ in the time interval $\mathcal{T} = [0,\tau)$, where $\tau$ is fixed. The unitary evolution of the system is determined by the  operator
\begin{equation}
    U(\tau, 0) = T e^{-i \int_0^\tau dt H(t)},  
\end{equation}
where $T$ is the time-ordering (chronological) operator.   Such an evolution  can be approximated by $N$ step-like Hamiltonian $H_N(t)$  consisting of partially constant Hamiltonians $H_1, H_2, \dots, H_N$ in equal time intervals $\mathcal{T}_k = [(k-1)\tau/N, k\tau/N)$ of length $\delta t = \tau / N$ 
(see Fig. 1). 
For any time $t\in \mathcal{T} = \cup_{k=1}^{N} \mathcal{T}_k$ we define 
\begin{eqnarray} \label{hn}
H_N(t) = H_k \ \text{for} \ t \in \mathcal{T}_k ,
\end{eqnarray}
where 
\begin{eqnarray} \label{pod}
 H_k =H(t_k), \quad  t_k=\frac{k-1}{N} \tau, \quad k=1, 2, ..., N. 
\end{eqnarray}
The corresponding evolution operator takes the form  
\begin{equation} \label{evo}
    U_N(\tau, 0) = T e^{-i \int_0^\tau dt H_N(t)}  = \prod_{k=1}^N e^{-\frac{i}{N} H_k \tau}  
\end{equation}
The last equality follows from the composition property  
\begin{eqnarray}
U(\tau, 0)=  U(t_{N+1}, t_{N}) U(t_{N}, t_{N-1})\dots U(t_3, t_2)\nonumber \\
 \times U(t_2, t_1)
\end{eqnarray}
 and  means that the evolution operator with burdensome time-ordering reduces to the product of  unitary operators generated by time-independent Hamiltonians.

In such an approach, the exact starting Hamiltonian $H(t)$ is the limit of the sequence $H_N$, i.e., 
\begin{eqnarray} \label{crqham}
    H(t) = \lim_{N \to \infty} H_N(t),
\end{eqnarray}
and as a consequence 
\begin{equation}
    U(\tau, 0) = \lim_{N \to \infty} U_N(\tau, 0).
\end{equation}

Let us notice that any such a step--like evolution can be described by an effective Hamiltonian $\tilde H_N$ which satisfies the relation
\begin{equation} \label{effuni}
    U_N(\tau, 0) = \exp[-i \tau \tilde H_N ] . \,
\end{equation}
We should have in mind that $\tau$ is fixed. If $\tau$ is changed to another value then  the effective Hamiltonian $\tilde H_N$ also changes accordingly.  

\subsection{Stochastic case}

Now,  let us consider a probabilistic case for which a system is driven by the time-dependent random Hamiltonian $H(t)$ in the time interval $\mathcal{T} = [0,\tau)$ with  fixed $\tau$.  A definition of a stochastic step--like driving $H_N(t)$ of a quantum system is analogous to a set of statistically independent Hamiltonians $\mathcal{H}_N=(H_1, H_2, \dots, H_N)$  which are random matrices  of  a joint matrix-pdf $\rho(\mathcal{H}_N)$ [cf. Eq. (\ref{pdf})]. Moreover, since all the  matrices in $\mathcal{H}_N$ are assumed to be statistically independent, for the time-dependent and random driving $H_N(t)$, 
one can postulate just a time--dependent matrix-pdf $\varrho_t(H)$ defined on a time interval $\mathcal{T}$ in such a way that 
 \begin{eqnarray}   \label{ro(t)}
\text{Pr}(H(t) \in D) = \int_D dH \varrho_t(H).  
\end{eqnarray}
One can represent the distribution $\rho(\mathcal{H}_N)$ in a time domain as
\begin{equation}
    \rho(\mathcal{H}_N) = \prod_{k=1}^N \varrho_{t_k} (H_k), \quad t_k=\frac{k-1}{N}\tau. 
\end{equation}

The case of a finite number of quenches $N$, when the evolution is driven by the Hamiltonian $H_N(t)$ in Eq. (\ref{hn}), is hereinafter referred to as a \emph{multiple random quench} (MRQ), whereas the limiting case for the Hamiltonian $H(t)$ [Eq. (\ref{crqham})] will be called as a \emph{continuous random quench} (CRQ). This continuous limit inherits  the condition that for any $t,s \in \mathcal{T}$ the Hamiltonians $H(t)$ and $H(s)$ are statistically independent random matrices. Notice that all  protocols for an arbitrary number $N$ (including limiting CRQ case) can be completely specified by the time-dependent pdf $\varrho_t(H)$.

\subsection{Effective Hamiltonian}

The effective Hamiltonian defined in Eq. (\ref{effuni}) can explicitly be obtained by using the relation (\ref{evo}) from which it follows that  
\begin{equation} \label{effuni1}
  \mbox{e}^{-i \tau \tilde H_N} = \prod_{k=1}^N \mbox{e}^{-\frac{i}{N} H_k \tau}  . \,
\end{equation}
For a given set  $\mathcal{H}_N=(H_1, H_2, \dots, H_N)$,  
we can calculate the effective Hamiltonian by use of  the Baker-Campbell- Hausdorff  formula \cite{hall}  for the operators  $A_k=-iH_k \tau$, namely,  
\begin{equation}
\prod_{k=1}^N e^{\frac{1}{N}A_k} = e^{Z_N}, 
\end{equation}
where  $Z_N$ has the following structure:  
\begin{eqnarray} \label{dynkin}
Z_N = \frac{1}{N}\sum_{i_1=1}^N A_{i_1} + \frac{1}{N^2}\sum_{i_1,i_2=1}^N \alpha_{i_1,i_2} [A_{i_1}, A_{i_2}] \nonumber\\
+ \frac{1}{N^3}\sum_{i_1,i_2,i_3=1}^N \alpha_{i_1,i_2,i_3} [A_{i_1}, [A_{i_2}, A_{i_3}]] + \dots
\end{eqnarray}
The parameters  $\alpha_{i_1,\dots,i_k}$ for $k = 2, 3, \dots$
can in principle be computed. Some effective algorithms for numerical calculations are presented e.g. in Refs. \cite{conv,casas}. However, the explicit form of the higher order terms is not straightforward since they involve more general nested commutators, like the commutators $[[A,B],[C,D]]$. They are not present in Dynkin's expansion \cite{dynkin} for two exponentials, as the commutators in Dynkin's form are ``segregated to the right'', but nevertheless the expansion  has a structure of Lie polynomials, i.e. it consists of commutators multiplied by numbers, which is crucial for the derivation of part of our results presented in this paper.

From now  on we will use an equivalent form of equation (\ref{dynkin}) given by the expansion of the commutators
\begin{eqnarray}\label{ass}
Z_N =   \frac{1}{N}\sum_{i_1=1}^N A_{i_1} + \frac{1}{N^2}\sum_{i_1,i_2=1}^N \beta_{i_1,i_2} A_{i_1} A_{i_2} \nonumber\\
+ \frac{1}{N^3}\sum_{i_1,i_2,i_3=1}^N \beta_{i_1,i_2,i_3} A_{i_1} A_{i_2} A_{i_3} + \dots
\end{eqnarray}
with a new set of coefficients $\beta_{i_1,\dots,i_n}$ which can be expressed by the $\alpha$-coefficients in (\ref{dynkin}). 

From the above relations (\ref{effuni1}) - (\ref{ass}) it follows that the effective Hamiltonian  $\tilde H_N$ can be represented by a series of polynomials $P_n(\mathcal{H}_N)$ of $n$-th degree of non--commuting matrix variables, namely, 
\begin{eqnarray} \label{series}
\tilde H_N = \frac{i}{\tau} \sum_{n=1}^\infty P_n(\mathcal{H}_N), 
\end{eqnarray}
where according to Eq. (\ref{ass}) one gets
\begin{eqnarray} \label{firstterms}
    P_1 (\mathcal{H}_N) &=& \frac{-i \tau}{N} \sum_{i_1=1}^N  H_{i_1},  \\
    P_2 (\mathcal{H}_N) &=&  \frac{(-i\tau)^2}{N^2} \sum_{i_1,i_2=1}^N \beta_{i_1,i_2} H_{i_1} H_{i_2},  \\
    P_3 (\mathcal{H}_N) &=& \frac{(-i\tau)^3}{N^3} \sum_{i_1,i_2,i_3 = 1}^N \beta_{i_1,i_2,i_3} H_{i_1} H_{i_2} H_{i_3},  
\end{eqnarray}
and so on. Although  an effective Hamiltonian $\tilde H_{N}$ obeying  (\ref{effuni}) always exists, the representation  (\ref{series}) is valid locally in some convergence domain of the series. There are various quantifiers estimating  the convergence domain  \cite{conv, conv1, conv2}.  However, the generalized case for $N$ exponentials requires a separate treatment (see Appendix A).  
It is crucial for our further reasoning to represent  the mean value of the effective Hamiltonian as
\begin{eqnarray} \label{avgseries}
\braket{\tilde H_N} = \frac{i}{\tau} \sum_{n=1}^\infty \braket{P_n(\mathcal{H}_N)} \,
\end{eqnarray}
and the variance--matrix as the series:
\begin{eqnarray} \label{varseries}
\mathrm{Var}(\tilde H_N)=  \sum_{n,m=1}^\infty S_{n,m} (\mathcal{H}_N) 
\end{eqnarray}
where 
\begin{eqnarray} \label{varelement}
[S_{n,m} (\mathcal{H}_N)]_{\alpha \beta} &=& \braket{[P_n(\mathcal{H}_N)]_{\alpha \beta}[P_m(\mathcal{H}_N)]^*_{\alpha \beta}} \nonumber \\ &-&\braket{[P_n(\mathcal{H}_N)]_{\alpha \beta}} \braket{[P_m(\mathcal{H}_N)]_{\alpha \beta}}^*. \,
\end{eqnarray}
To keep  mathematical rigour and to ensure the existence of these averages one can simply assume that the ensemble is contained in the convergence domain.

\subsection{Commutation in the statistical sense}

At the end of this introductory part we define a special condition required in the following proofs that we call commutation  in the statistical sense. To this aim,  let us notice that the mean value of the Hamiltonian $H(t)$ for the CRQ protocol  can be  expressed as an ensemble average over the distribution $\varrho_t(H)$, namely, 
\begin{equation}
\braket{H(t)} = \int_\mathcal{P} dH H \varrho_t(H) . \,
\end{equation}
We say that two observables $O_1$ and $O_2$ commute in the statistical sense if their mean values with respect to the matrix ensemble commute, i.e. $[\braket{O_1}, \braket{O_2}] = 0$. In our particular case, we say that the whole MRQ protocol, defined solely by the distribution $\varrho_t(H)$, commutes in the statistical sense if
\begin{eqnarray} \label{comcon}
[\braket{H(t)}, \braket{H(s)}] = 0
\end{eqnarray}
holds true for any $t,s \in \mathcal{T}$. Notice that the above condition also implies $[\braket{H_N(t)}, \braket{H_N(s)}] = 0$ for an arbitrary number of quenches $N$.
We stress that commutation in the statistical sense is a weaker condition than standard commutation.  In particular, it means that  the first moments in different time instances can be simultaneously diagonalized.  

\section{Self--averaging limit}

In this section we present two propositions implying explicit conditions for self--averaging,  i.e. when a deterministic description can effectively approximate  an essentially random system. 

In the following we will use the abbreviation 
\begin{equation}
\beta(n) = \max_{i_1,i_2,\dots,i_n} | \beta_{i_1 \dots i_n} |.
\end{equation}
and a dimensionless quantity 
\begin{equation}
    K(n) = \max_{t \in \mathcal{T}} \int_\mathcal{P} dH \tau^n \| H \|^n \varrho_t(H).
\end{equation}
In order to simplify notation we  also use the notation: $K(n,m) \equiv K(n+m) + K(n)K(m)$.

Now, we can state our main result. Let $\mathcal{H}_N$ be a set of random matrices representing a stochastically controlled quantum system via the MRQ protocol. The mean values of polynomials $P_n(\mathcal{H}_N)$ and $S_{n,m} (\mathcal{H}_N)$, which constitute the expansion of the mean effective Hamiltonian $\braket{\tilde H_N}$ in Eq.(\ref{avgseries}) and $\mathrm{Var}(\tilde H_N)$ in Eq.(\ref{varseries}),  respectively, satisfy the following conditions:  
\begin{theorem} \label{theorem2}
For the MRQ protocol in the time interval $\mathcal{T} = [0,\tau)$ with pdf $\varrho_t(H)$,  
\begin{equation}
\| S_{n,m}(\mathcal{H}_N) \| \le R_{n+m} (N) \beta(n)\beta(m) K(n,m). \,
\end{equation}
for any $n+m<N$. Moreover, if MRQ commutes in the statistical sense [Eq. (\ref{comcon})], then for $N>n>1$:
\begin{equation}
\| \braket{P_n(\mathcal{H}_N)} \| \le R_{n}(N) \beta (n) K(n), 
\end{equation}
where 
\begin{equation} \label{poly}
R_{n} (N) =1 - \frac{N!}{N^{n}(N-n)!} =  O(\frac{1}{N}). \,
\end{equation}
\end{theorem}
\begin{theorem} \label{theorem1}
For the MRQ protocol in the time interval $\mathcal{T} = [0,\tau)$ with time-independent distribution $\varrho(H) \equiv \varrho_t(H) = \varrho_s(H)$ for any $t,s \in \mathcal{T}$, the following
\begin{eqnarray}
    \braket{P_{2n}(\mathcal{H}_N)} = S_{2n, 2m+1}(\mathcal{H}_N) = S_{2n+1, 2m}(\mathcal{H}_N) = 0
\end{eqnarray}
holds true for any $n,m \in \mathbb{N}$. In addition, if $\varrho(-H) =  \varrho(H)$, then $\braket{P_{n}(\mathcal{H}_N)} = 0$.
\end{theorem}
Note that a rough estimate of $\beta(n)$ shows that $\beta(n) < 1$ and that it is a decreasing function of $n$. Further, if one additionally assumes a convergence condition to be satisfied,  $K(n)$ is an exponentially decreasing function of $n$. Hence, one concludes that for Hamiltonians and time scales satisfying convergence,  the expected value and variance of the effective Hamiltonian can be approximated by a finite number of terms in Eq. (\ref{series})  which for large $N$ decrease as $O(1/N)$.  

\subsection{Variance of the unitary time-evolution}
For the MRQ protocols satisfying convergence condition (i.e. for low driving frequencies or short time scales) the variance-matrix satisfies 
\begin{eqnarray}
\| \mathrm{Var}(\tilde H_N) \|  \le \sum_{n,m=1}^\infty \| S_{n,m} (\mathcal{H}_N) \| = O(\frac{1}{N}) . \,
\end{eqnarray}
This condition is sufficient to show that not only the variance of the time evolution unitary matrix  decreases as $O(1/N)$ but also an arbitrary product of such  matrices decreases as $O(1/N)$ (see Appendix D and F). In other words a vanishing variance of the local generator (i.e. the effective Hamiltonian) implies the vanishing of the variance of the unitary matrix in a larger domain. Thus, for an arbitrary distribution $\rho_t(H)$ supported on a bounded interval  
\begin{eqnarray}\label{sn}
S_N:=\| \mathrm{Var}[U_N (\tau,0)] \| = O(\frac{1}{N}) . \,
\end{eqnarray}
One infers that the magnitude of the variance-matrix for  a convergent series (\ref{varseries}) becomes  arbitrarily small with an increasing  number $N$ of quenches,  i.e. for  the step--like MRQ protocol one obtains self-averaging of the time evolution of the quantum system.   In particular, in the limiting case of  control given by the CRQ protocol one obtains
\begin{eqnarray} \label{varlimit}
\lim_{N \to \infty} S_N &=& 0
\end{eqnarray}
and this implies that $U_N$ (or correspondingly $\tilde H_N$) approaches a degenerate random variable,  i.e. it converges almost surely to its mean value, 
\begin{eqnarray}
\lim_{N \to \infty} T e^{-i \int_0^\tau dt H_N(t)} = \lim_{N \to \infty} \braket{T e^{-i \int_0^\tau dt H_N(t)}}. \,
\end{eqnarray}

This result can be of interest for potential experimental applications. For  general MRQ protocols one expects that the driving of a quantum system given by step--like independent random changes of the Hamiltonian becomes more regular in   the limiting control of the CRQ protocol which can serve as an effective classification scheme of different stochastic time evolutions  (convergent to the same self-averaged one).

Notice that commonly defined self-averaging condition (\ref{1}) involves scaling of the variance by the square of the mean value. {\it Per analogiam}, we can scale the quantity $S_N$ by the quantity $\| \braket{U_N(\tau,0)} \|^2$, however, any unitary matrix is constant in the Frobenius norm (\ref{norm}) and equal to the dimension $M$ of the Hilbert space  [see Eq. (\ref{norm})], thus
\begin{equation}
\frac{\| \mathrm{Var}[U_N (\tau,0)] \|}{\| \braket{U_N(\tau,0)} \|^2} =\frac{1}{M} S_N = O(\frac{1}{N}) . \,
\end{equation}
and further on we will just use the $S_N$.


\subsection{Average of the unitary time-evolution}
Let us now examine the mean value of the effective Hamiltonian. From the second part of Proposition 1, if an additional assumption of  commutation in the statistical sense holds true [cf. Eq. (\ref{comcon})], a growing number of quenches $N$ results in decreasing the absolute value of the non-commutative part of the  series  (\ref{series}). In particular, it implies that 
\begin{equation} \label{limitmean}
\|\braket{\tilde H_N} - \frac{1}{\tau} \int_0^\tau dt \braket{H_N(t)} \| = O(\frac{1}{N}).
\end{equation}
This condition is sufficient  to derive an analogous relation in terms of the unitary operator:
\begin{equation} \label{dn}
D_N := \|\braket{T e^{-i \int_0^\tau dt H_N(t)}} - T e^{-i \int_0^\tau dt \braket{H_N(t)}} \| = O(\frac{1}{N}). 
\end{equation}
which is valid for an  arbitrary distribution $\rho_t(H)$ supported on the bounded interval (see Appendix E and F). In the limit it  gives
\begin{equation} \label{evocom}
\lim_{N \to \infty} \braket{T e^{-i \int_0^\tau dt H_N(t)}} = T e^{-i \int_0^\tau dt \braket{H(t)}} \, .
\end{equation}
Notice that in fact, due to the condition (\ref{comcon}) the  time ordering can be dropped here, however,  we left it since in the next section we numerically compute the quantity $D_N$ in a more general case.

Surprisingly, the numerical simulation performed for  qubits and presented in the next Section confirms the validity of the formula (\ref{dn}) also in the non-commuting case. This  observation, although very particular,  suggests the conjecture that  Eq. (\ref{dn}) can be valid generally and thus can be successfully applied in practice as an extremely useful tool simplifying very complicated calculations of averaged unitary evolutions.

A special case of Hamiltonians commuting in the statistical sense are exemplified by  protocols with time-independent distributions such that $\varrho_t(H) = \varrho_s(H) \equiv \varrho(H)$. In such a case the set of $\mathcal{H}_N$  consists of independent and identically distributed random matrices and we refer this protocol as IID protocol. Upon Proposition 2  we conclude that only odd terms contribute to the series, 
\begin{equation}
\braket{\tilde H_N} = \frac{i}{\tau} \sum_{n=0}^\infty \braket{P_{2n+1}(\mathcal{H}_N)}, \,
\end{equation}
 Moreover, the second part of Proposition 2 also implies that for an even pdf $\braket{\tilde H_N}=0$ or equivalently $\braket{U_N(\tau,0)}= \mathbb{1}$ for an arbitrary number of quenches $N$. Also one half of the terms of the  series (\ref{series}) vanish and the variance-matrix Eq.(\ref{varseries}) reduces to the series
\begin{equation}
\mathrm{Var}(\tilde H_N) = \sum_{n,m=0}^\infty \braket{S_{2n,2m} (\mathcal{H}_N) + S_{2n+1,2m+1} (\mathcal{H}_N)}. \,
\end{equation}
For this special case the instantaneous first moment $\braket{H(t)}$ is time-independent and equal to the effective Hamiltonian in the CRQ limit, i.e., 
\begin{eqnarray}
\braket{H(t)} = \lim_{N \to \infty} \tilde H_N  = \int_\mathcal{P} dH H \varrho(H).
\end{eqnarray}


\section{Numerical treatment}
In this section we numerically analyze MRQ protocol applied to a two-dimensional Hilbert space which describes quantum two-level systems. 
Despite their simplicity, two-level-systems play a crucial role in many branches of theoretical and applied physics.
The celebrated NMR (nuclear magnetic resonance) is probably the most spectacular example which is one of the primary stages for dynamical decoupling and averaging schemes \cite{haber1,haber2}.  Our work, at least partially, goes beyond that studies since here we apply averaging to stochastically driven two--level systems which, however,  can mimic the realistic but randomly disturbed NMR systems and hence can be be of potential applicability not only in theoretical studies of quantum random dynamics but also in magnetic--based imaging ranging from solid state physics, via quantum chemistry up to medical physics. 
Moreover, two--level systems, the qubits per se, are the basic building blocks for encoding quantum information. Unfortunately, decoherence and uncontrollable fluctuations (both deterministic and random)  seem to be one of several obstructions for an effective implementations of the power of quantum information processing and quantum computing. Stochastic averaging is one of potential candidates for controlling and  correcting errors of a certain type. 

For a random evolution of a qubit we represent time-dependent Hamiltonian in the form 
\begin{equation} \label{qubitham}
H(t) = \frac{1}{2} \vec \alpha(t) \cdot \vec \sigma,  
\end{equation}
where $\vec \sigma$ is a vector of Pauli matrices and  $\vec \alpha(t)$ is a  vector of independent random components distributed according to the normal distribution with mean values $\mu_i(t)$ (for $i=1,2,3$) and the same variance 
 $\sigma^2$ for all components. 

We consider its three protocols: (i) the time-independent IID protocol, (ii) the  time-dependent commuting and (iii) non-commuting cases. For these three cases, we calculate $S_N$ and $D_N$  with respect to a number of quenches $N$. For the IID protocol  
\begin{eqnarray}\label{IID}
\vec \mu_{I}(t) = \vec \mu
\end{eqnarray}
with magnitude $|\vec \mu| = \mu$. Notice that if $\mu = 0$ or if the vector $\vec \mu$ has  only one of three components non-vanishing, we obtain the Gaussian Unitary Ensemble \cite{mehta} for the  qubit space. For the time-dependent commuting case we take the  single harmonic 
\begin{eqnarray}\label{statcom}
\vec \mu_C(t) =  \mu \left( \sin(\omega t) , 0, 0 \right)
\end{eqnarray}
and for the non-commuting case we assume 
\begin{equation}\label{statnoncom}
    \vec \mu_N(t) = \mu \left( \sin(\omega t) , \cos(\omega t), 0 \right).
\end{equation}

Results presented in Figs.  2-4  reveal that the quantities $S_N$ and $D_N$  obey  power-law behaviour  $1/N$ for sufficiently large values of $N$. Notice that in Eqs. (\ref{sn}) and (\ref{dn}) we state only that they behave at least as $O(1/N)$.   
Surprisingly, the same behaviour is also observed for the non-commuting case. 
According to numerical simulations (Fig. 4) for this particular driving of the two-level system it is seen that the quantity $D_N$ vanishes as one over the number of quenches. This result suggests that Eq. (\ref{dn}) can be valid in a more general case. However,  this subject needs further studies for higher dimensional systems and other distributions of the MRQ protocol.  

\begin{figure}
    \centering
    \begin{minipage}[b]{0.2\textwidth}
        \includegraphics[width=0.85\textwidth, angle=270]{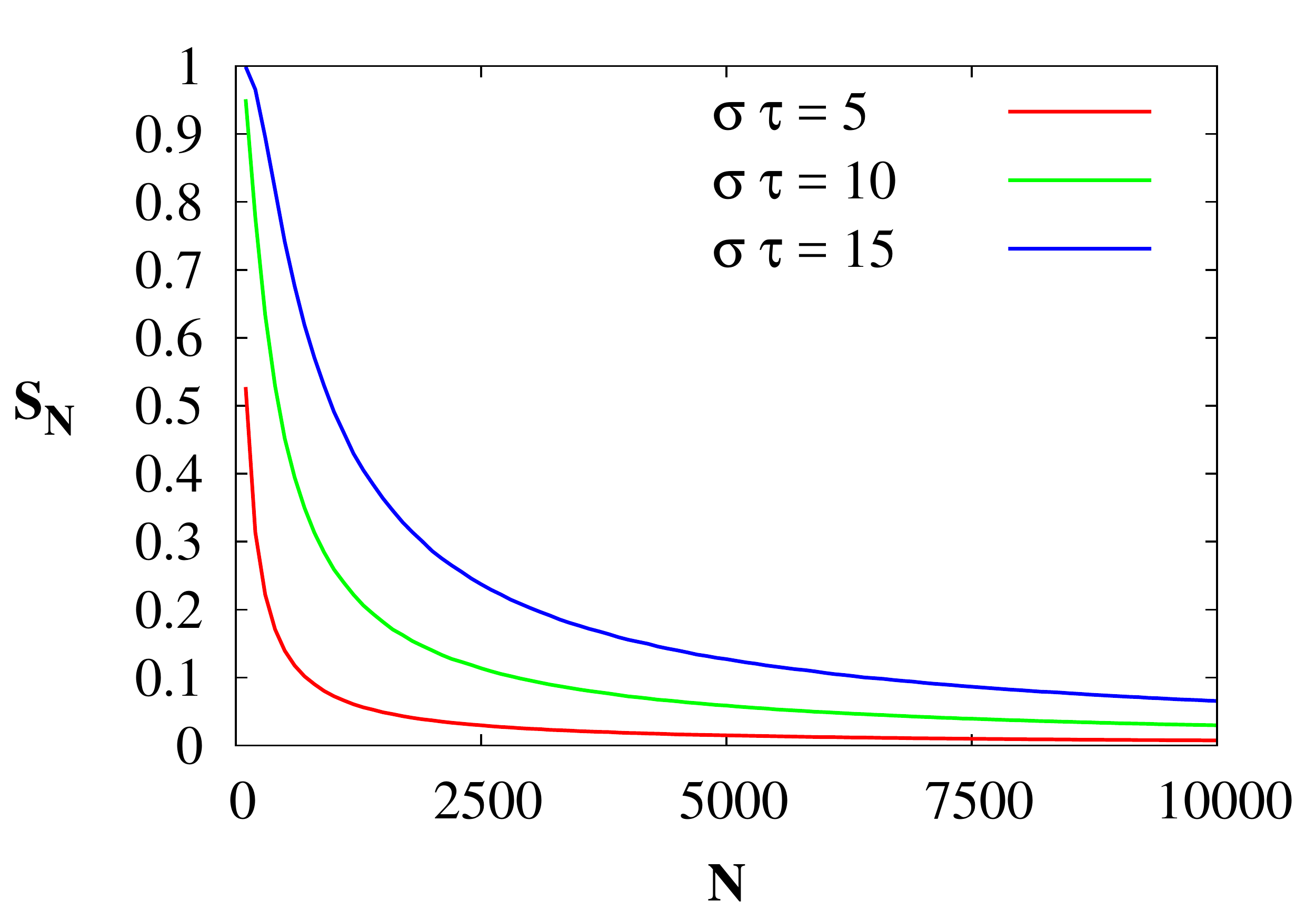}
    \end{minipage}
    \hfil
    \begin{minipage}[b]{0.2\textwidth}
        \includegraphics[width=0.85\textwidth, angle=270]{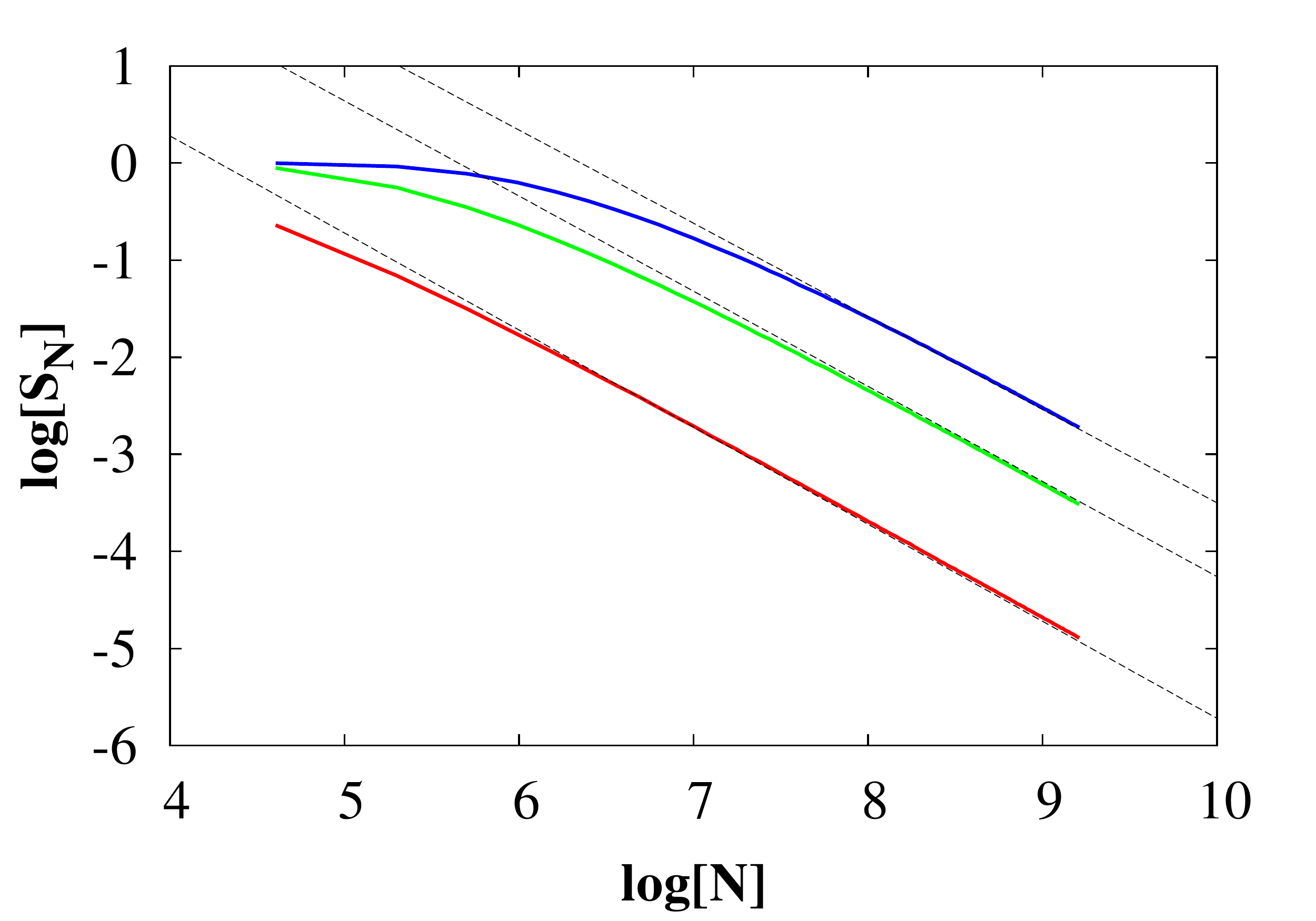}
    \end{minipage}
    \begin{minipage}[b]{0.2\textwidth}
        \includegraphics[width=0.85\textwidth, angle=270]{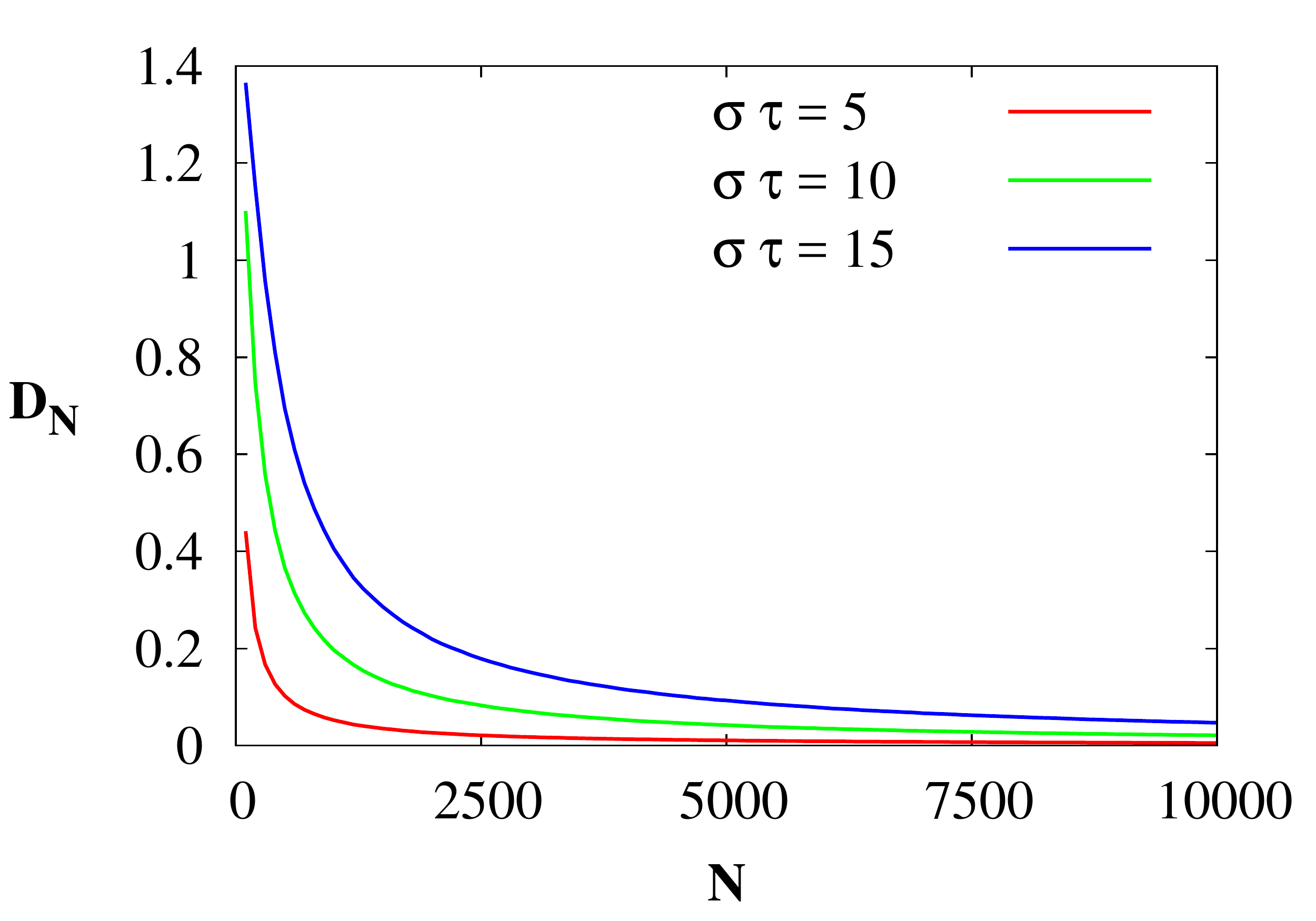}
    \end{minipage}
    \hfil
    \begin{minipage}[b]{0.2\textwidth}
        \includegraphics[width=0.85\textwidth, angle=270]{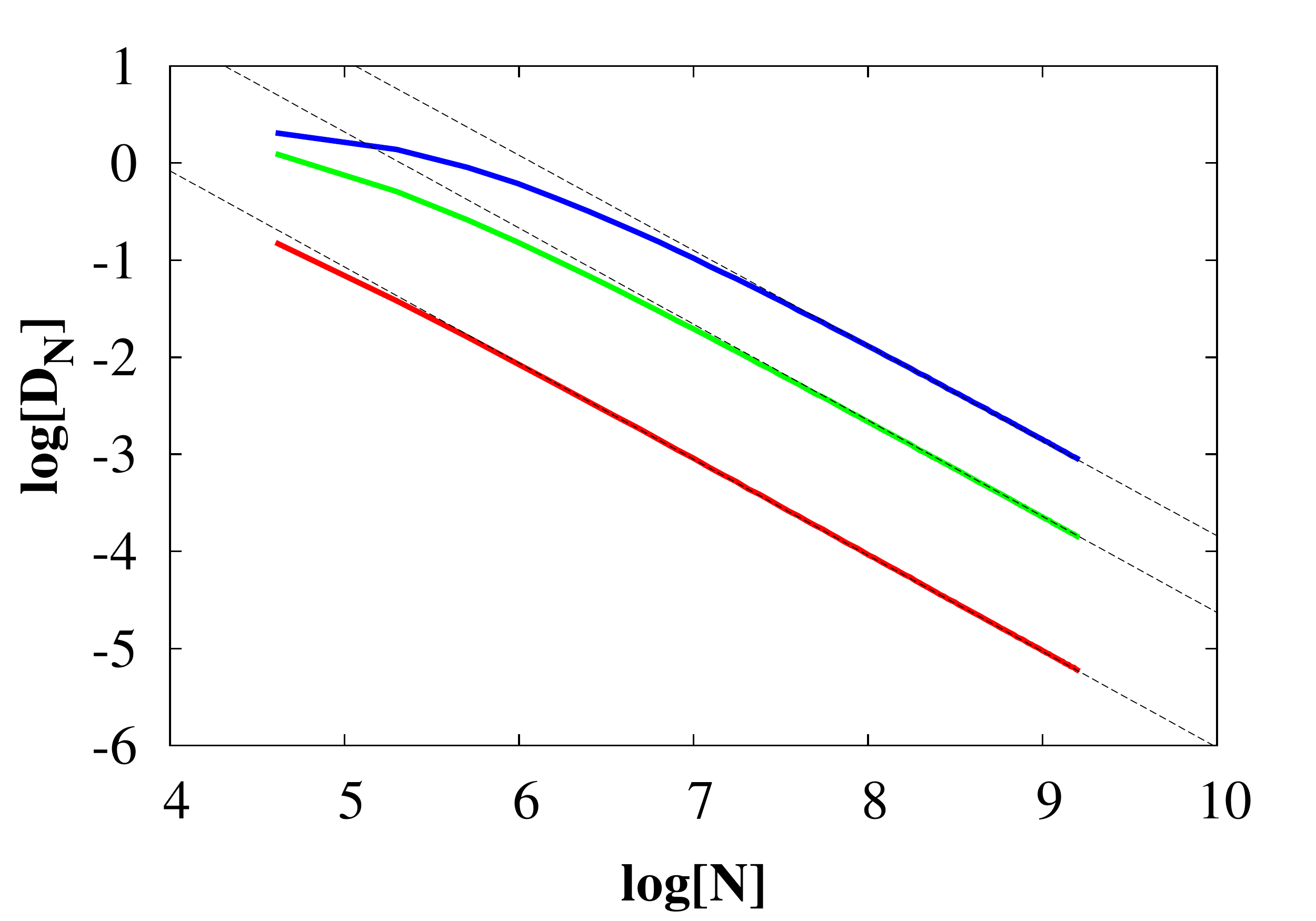}
    \end{minipage}
    \caption{Graphs of $S_N$ and $D_N$ with respect to the number of quenches $N$ for the IID protocol [Eq. (\ref{IID})] for different values of the  dimensionless parameter $\sigma \tau$ with the constant value $\mu \tau = 1$. Corresponding log-log graphs show apparently  a power law for large  values of $N$. From linear regression we obtain: $S_N~\propto~N^{-1.00}, N^{-0.98}, N^{-0.96}$ and $D_N~\propto~N^{-0.99}, N^{-0.99}, N^{-0.98}$ with respect to an increasing value of the parameter $\sigma \tau = 5, 10, 15$. }
    \label{figIID}
\end{figure}

\begin{figure}
    \centering
    \begin{minipage}[b]{0.2\textwidth}
        \includegraphics[width=0.85\textwidth, angle=270]{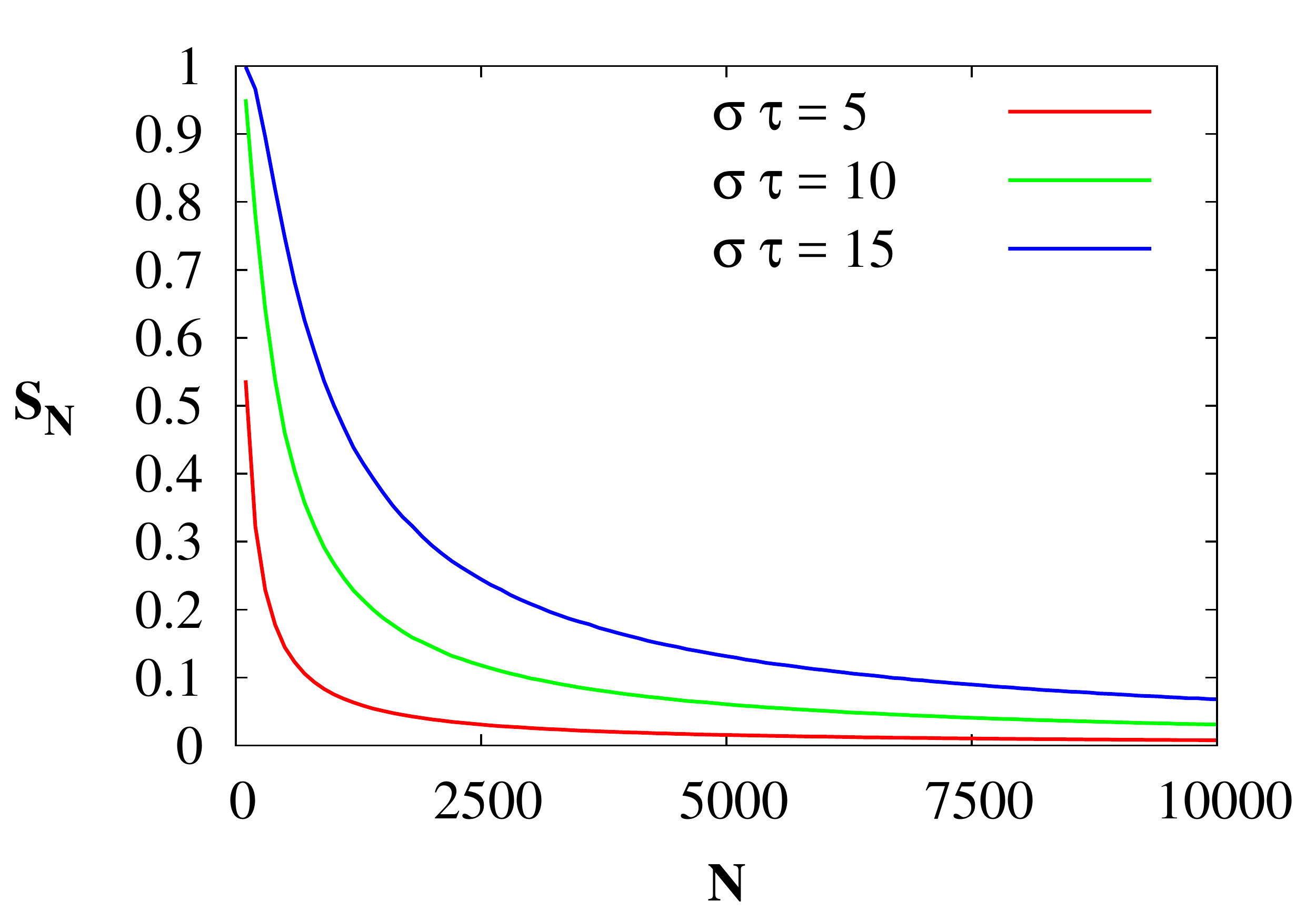}
    \end{minipage}
    \hfil
    \begin{minipage}[b]{0.2\textwidth}
        \includegraphics[width=0.85\textwidth, angle=270]{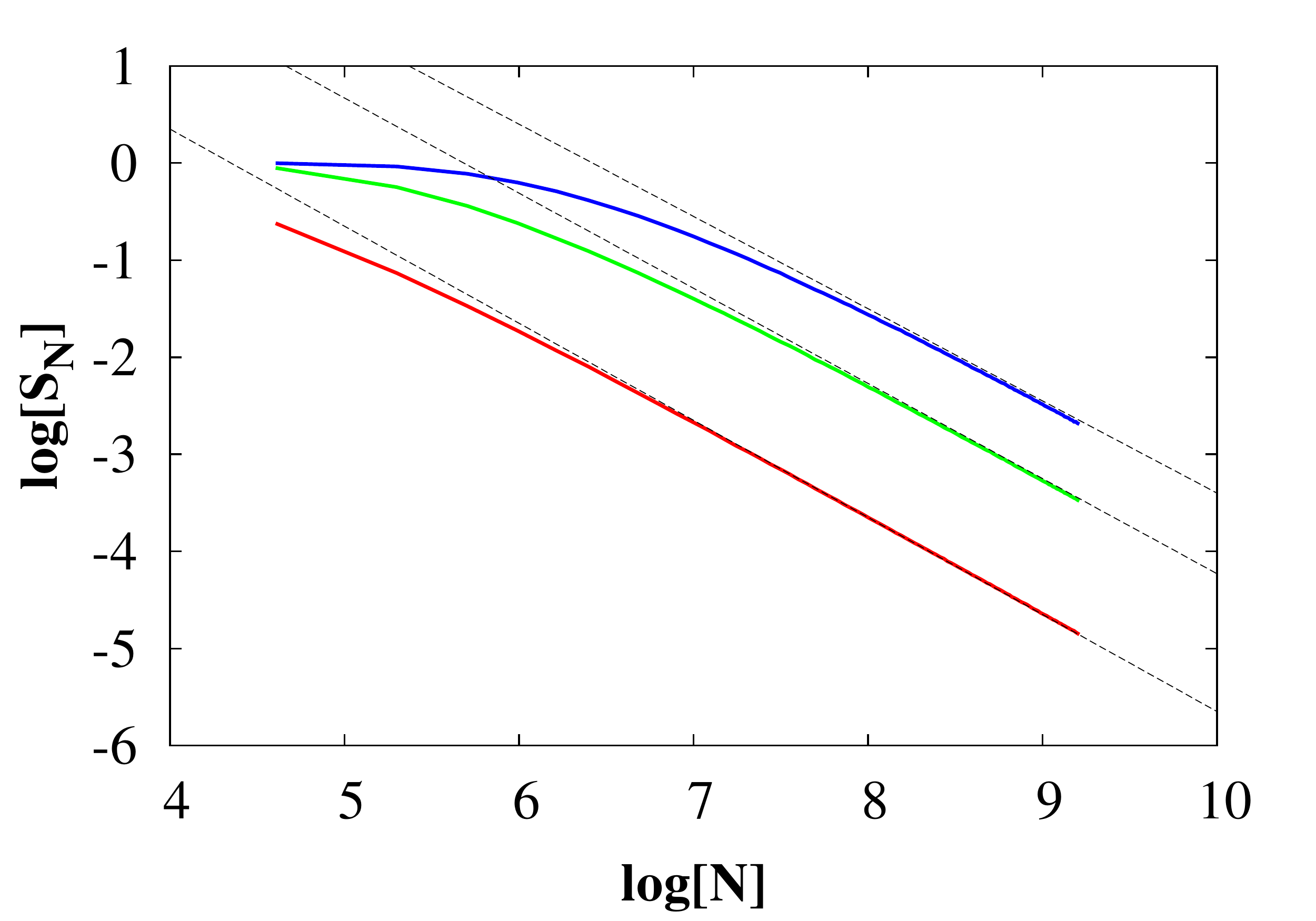}
    \end{minipage}
    \begin{minipage}[b]{0.2\textwidth}
        \includegraphics[width=0.85\textwidth, angle=270]{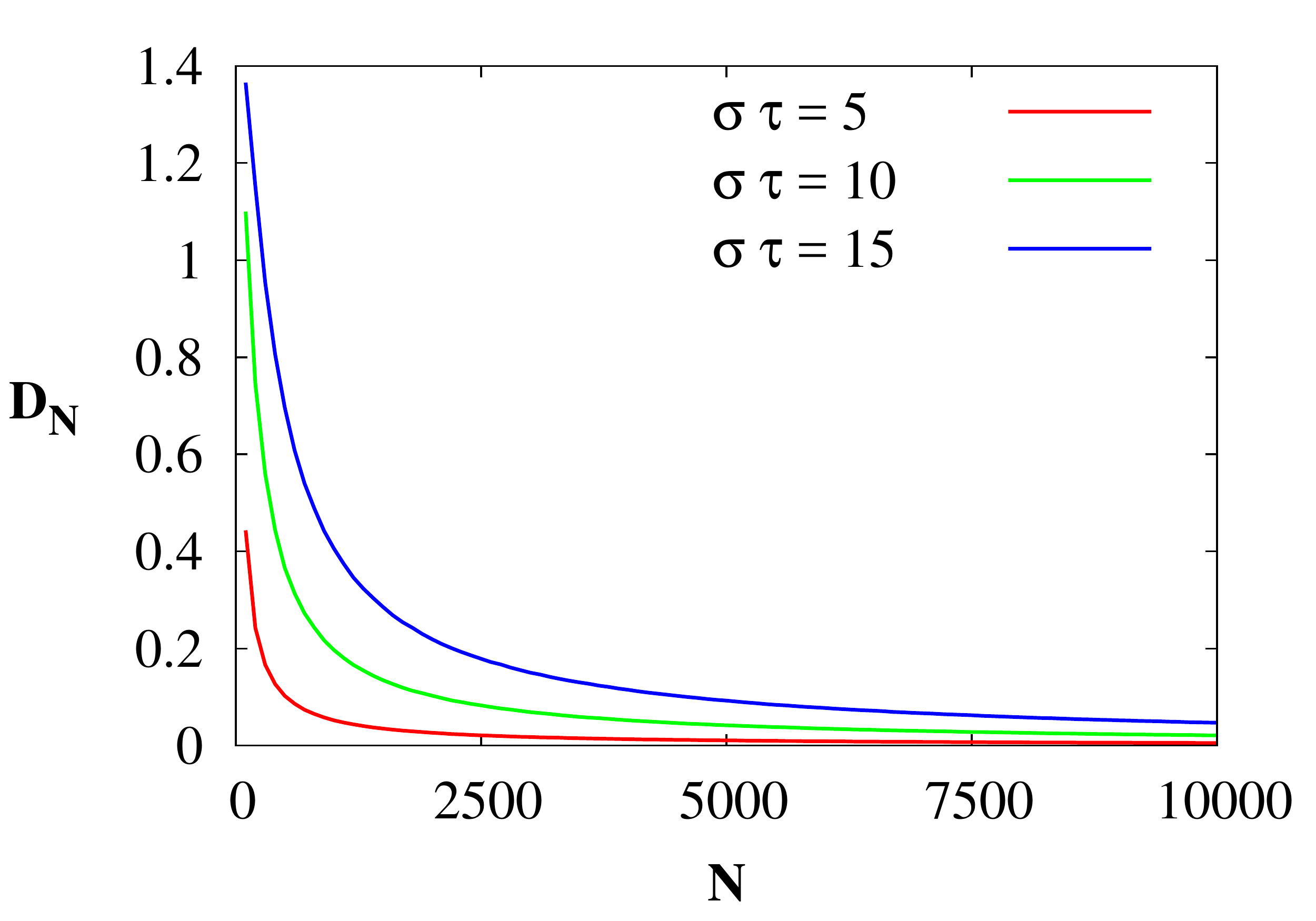}
    \end{minipage}
    \hfil
    \begin{minipage}[b]{0.2\textwidth}
        \includegraphics[width=0.85\textwidth, angle=270]{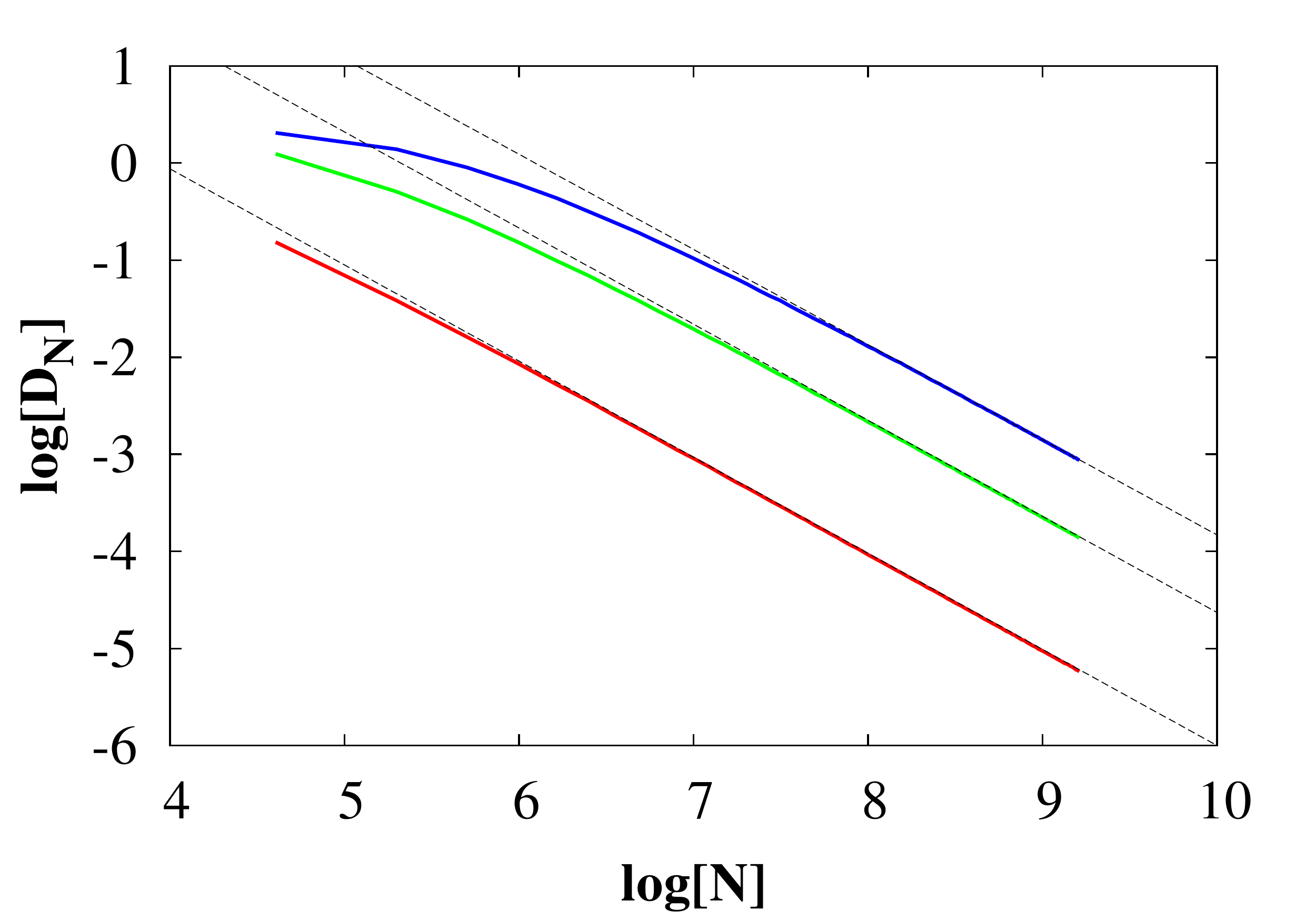}
    \end{minipage}
    \caption{Graphs of $S_N$ and $D_N$ with respect to the number of quenches $N$ for the statistically commuting case [Eq. (\ref{statcom})] and for selected  values of the dimensionless parameter $\sigma \tau$ with the constant value $\mu \tau = 1$ and $\omega \tau = \pi /4$. Corresponding log-log graphs show apparently a power law for large values of $N$. From linear regression we obtain: $S_N~\propto~N^{-1.00}, N^{-0.98}, N^{-0.95}$ and $D_N~\propto~N^{-0.99}, N^{-0.99}, N^{-0.98}$ with respect to an increasing value of the parameter $\sigma \tau = 5, 10, 15$. }
    \label{figCom}
\end{figure}

\begin{figure}
    \centering
    \begin{minipage}[b]{0.2\textwidth}
        \includegraphics[width=0.85\textwidth, angle=270]{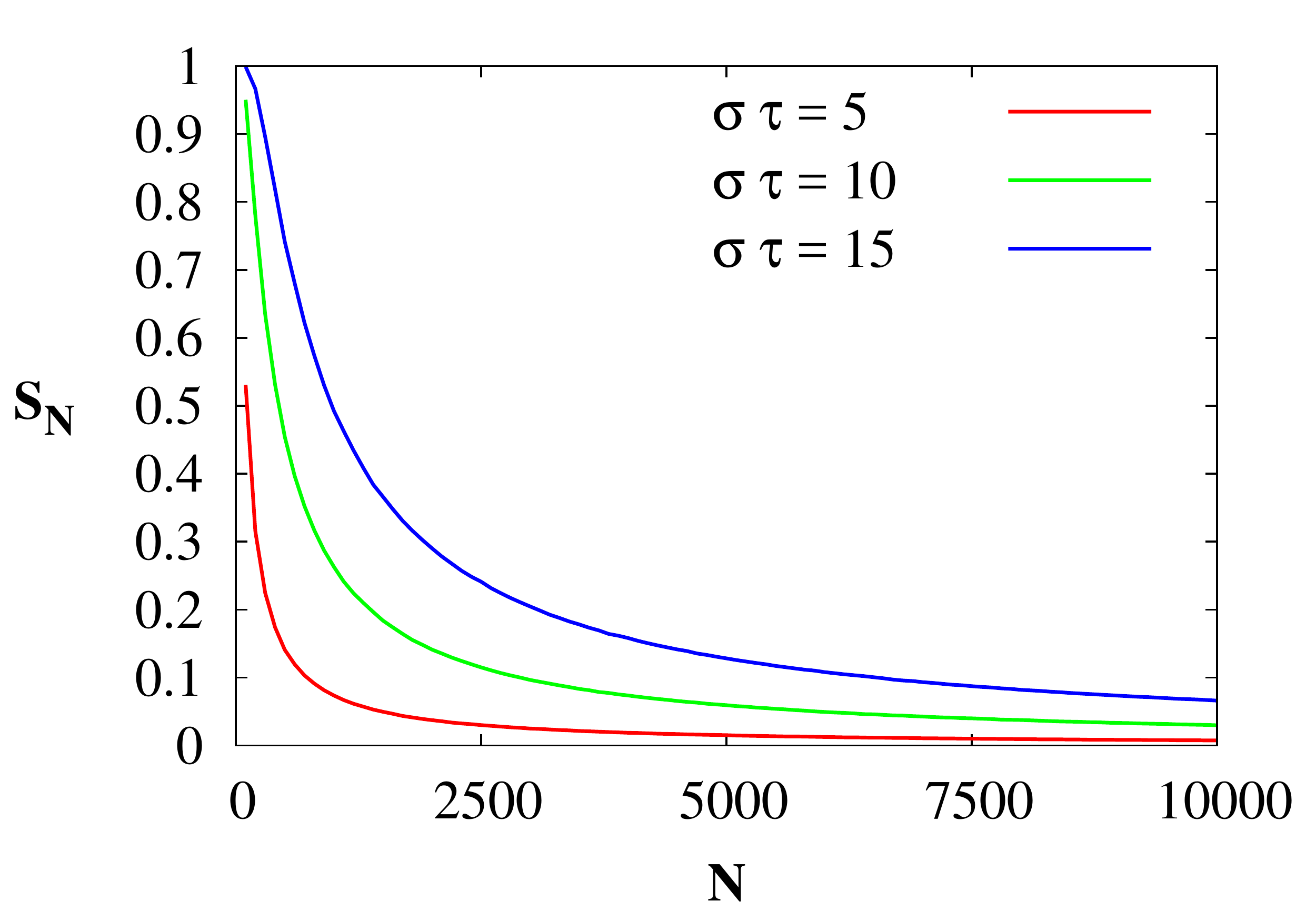}
    \end{minipage}
    \hfil
    \begin{minipage}[b]{0.2\textwidth}
        \includegraphics[width=0.85\textwidth, angle=270]{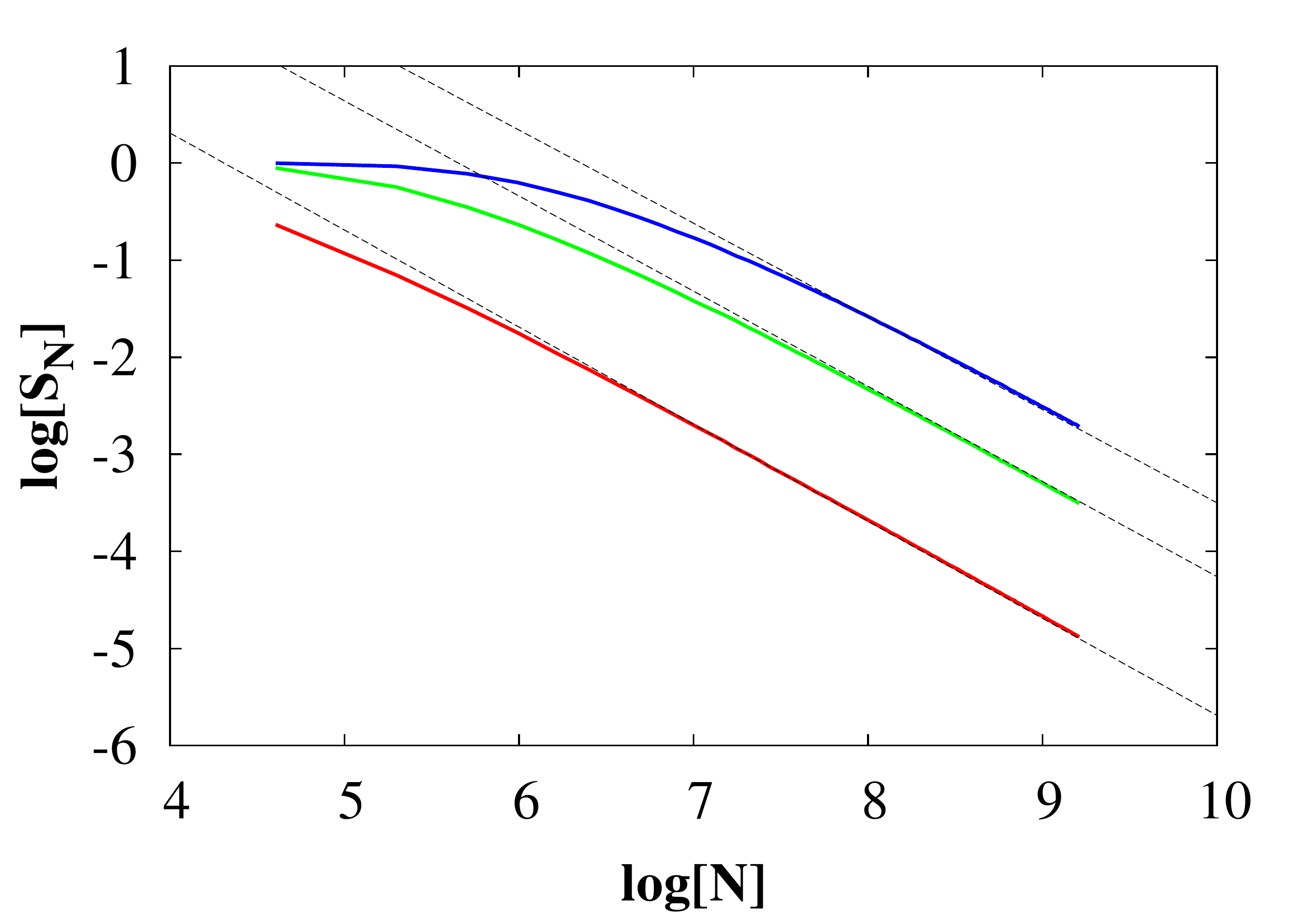}
    \end{minipage}
    \begin{minipage}[b]{0.2\textwidth}
        \includegraphics[width=0.85\textwidth, angle=270]{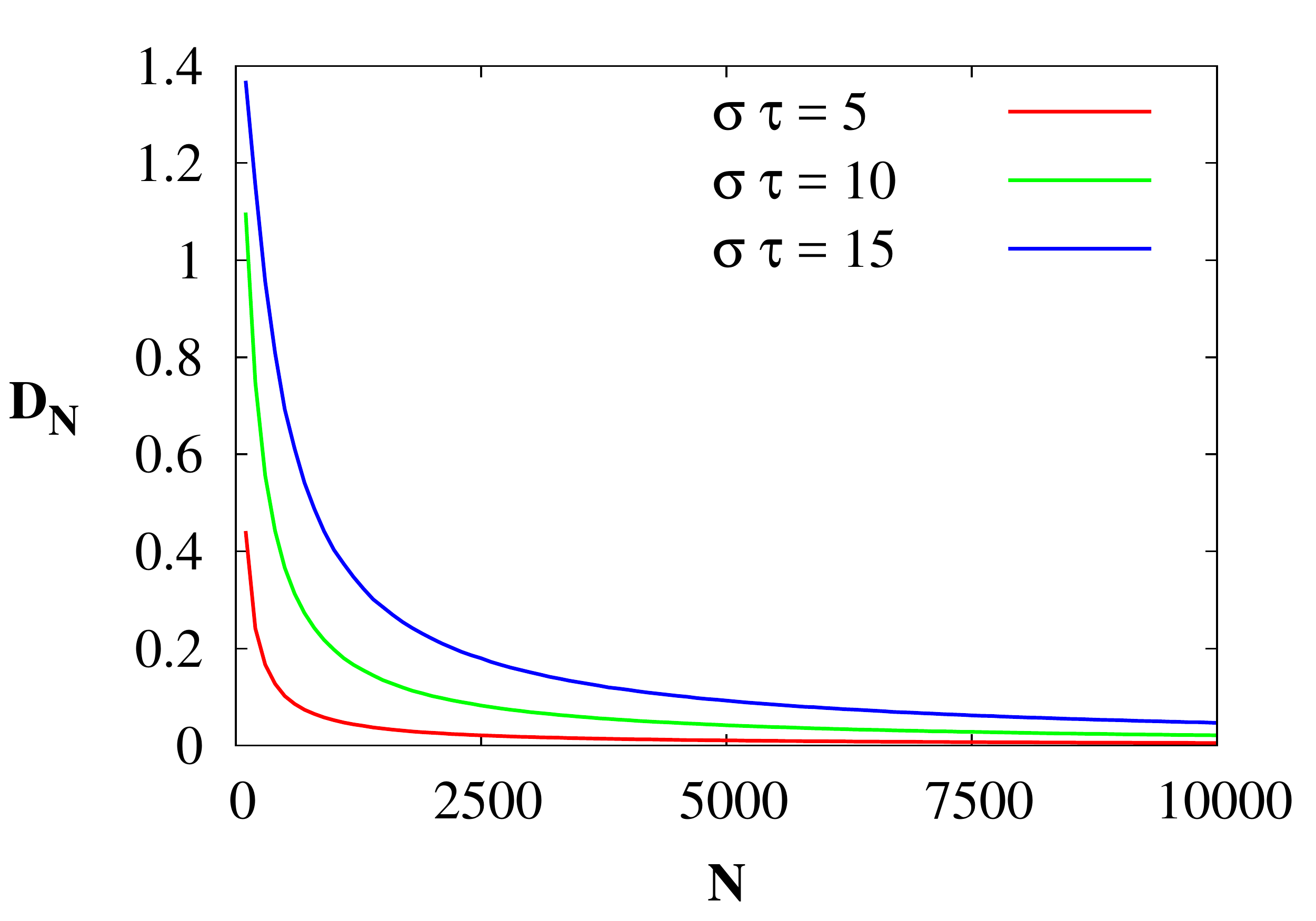}
    \end{minipage}
    \hfil
    \begin{minipage}[b]{0.2\textwidth}
        \includegraphics[width=0.85\textwidth, angle=270]{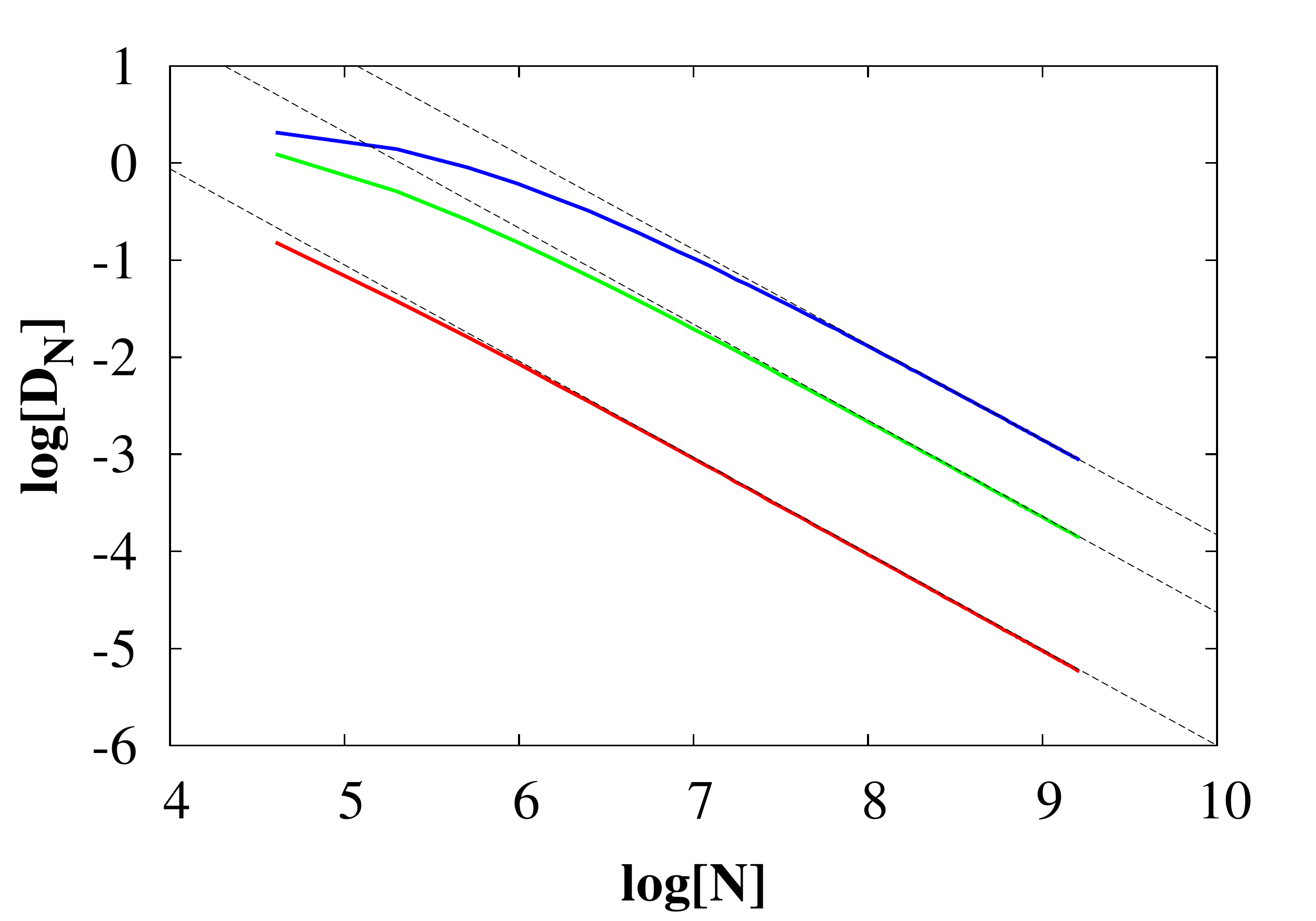}
    \end{minipage}
    \caption{Graphs of $S_N$ and $D_N$ with respect to the number of quenches $N$ for the statistically non-commuting case [Eq. (\ref{statnoncom})] and  for various values of the dimensionless parameter $\sigma \tau$ with the constant value $\mu \tau = 1$ and $\omega \tau = \pi /4$. Corresponding log-log graphs show apparently a power law for large values of $N$. From linear regression we obtain: $S_N~\propto~N^{-1.00}, N^{-0.98}, N^{-0.96}$ and $D_N~\propto~N^{-0.99}, N^{-0.99}, N^{-0.98}$ with respect to an increasing value of the parameter $\sigma \tau = 5, 10, 15$. }
    \label{figNonCom}
\end{figure}

\section{Summary} 

Time--dependent and stochastically driven quantum systems are very important for modern applications since they effectively mimic external control applied to gain desired dynamic properties. In particular, a stochastic description  becomes unavoidable either if there is a certain degree of uncertainty or randomness affecting the control strategy or if there is disorder essentially present in the system under consideration. An examples are random decoupling schemes for quantum dynamical control and error suppression \cite{santos}. 
In many cases, a proper description requires random operators~\cite{randy} resulting in the models yet very elegant and effective but not easy to analyze. That is why every result simplifying the analysis or serving as a useful tool is not only 'theoretically attractive' but also is of great practical importance. Self--averaging is one among such concepts developed to investigate a certain class of stochastically modified quantum systems which still remain challenging not only for mathematical physicists  (cf. Chapter 3. in Ref.\cite{randy}) but also for these who want effectively and credibly simulate quantum dynamics of non--trivial systems.      
In our work we formulated and studied quantum systems undergoing Multiple  Random Quench protocols. A more abstract approach on the problems studied in this paper can be found in Ref. \cite{hiller}. General results  put in the
framework of convolution semigroups are presented in the book \cite{gren}. Here, we investigated statistical properties of an effective unitary dynamics with an emphasis on the self--averaging property. We recognized that for a broad class of randomly driven systems satisfying relatively non--restrictive conditions the self--averaging phenomenon occurs and can be utilized for a considerable simplification of the treatment of such systems. Our findings, derived via mathematically rigorous reasoning, are supported by numerical calculations. Such a test allows not only to verify theoretical and more formal predictions but also helps to formulate a conjecture applicable  beyond mathematically proved cases.  Our result for a bridge between formal but sometimes highly restricted mathematical treatment and more informal purely numeric modeling applicable to a broad but not precisely defined class of random systems. We hope that our modest contribution -- despite of enhancing our understanding of quantum stochastic dynamics -- can also serve as a training ground suitable for testing numerical tools: even if one is interested in dynamic properties of  systems which essentially do not fulfill requirements of the propositions stated in this paper such that numerical treatment becomes unavoidable, one can still verify credibility of numerics applied to  a class of systems described here.

\appendix
\section{Estimation of the convergence domain}
Let us express the  polynomials $P_n(\mathcal{H}_N)$ in Eq. (\ref{series}) in the associative representation:
\begin{equation} \label{termavg}
P_n (\mathcal{H}_N) = \frac{(-i \tau)^n}{N^n} \sum_{i_1,\dots,i_n}^N \beta_{i_1,\dots,i_n} H_{i_1} H_{i_2} \dots H_{i_n}. \,
\end{equation}
From this relation one can make the estimate 
\begin{equation}
\begin{split}
    \sum_n \| P_n(\mathcal{H}_N) \| &\le \sum_n \frac{\tau^n}{N^n} \sum_{i_1,\dots,i_n}^N | \beta_{i_1,\dots,i_n} | \| H_{i_1} H_{i_2} \dots H_{i_n} \| \\
    & \le \sum_n \frac{\tau^n}{N^n} \sum_{i_1,\dots,i_n}^N \| H_{i_1} \| \| H_{i_2} \| \dots \| H_{i_n} \|
\end{split}
\end{equation}
Under the assumption that 
\begin{equation} \label{convcond}
    H_i \subset \mathcal{C} = \{ H : \|H\| < 1/\tau \}
\end{equation} 
for $i = 1,2,\dots,N$, we infer that  there exist a number $M < 1$ such that
\begin{equation}
\begin{split}
    \sum_n \| P_n(\mathcal{H}_N) \| & \le \sum_n \frac{1}{N^n} \sum_{i_1,\dots,i_n}^N M^n \\
    & \le \sum_n M^n < \infty
\end{split}
\end{equation}
and this proves the absolute convergence of the  series (\ref{series}) if $\mathcal{P} \subset \mathcal{C}$.

\section{Proof of Theorem 1}
\subsection{First statistical moment}
From the Lie representation of the polynomials $P_n (\mathcal{H}_N)$ we conclude that any of them can be represented by a linear combination of the following terms
\begin{equation}
H_{i_1} H_{i_2} \dots H_{i_{p-1}} [H_{i_p}, H_{i_{p+1}}] H_{i_{p+2}} \dots H_{i_{n}}
\end{equation}
where $n > p>1$. Thus, for a subset of different indices $i_1 \neq i_2 \neq \dots \neq i_n$ one  obtains
\begin{equation}
\begin{split}
&\braket{H_{i_1} H_{i_2} \dots H_{i_{p-1}} [H_{i_p}, H_{i_{p+1}}] H_{i_{p+2}} \dots H_{i_{n}}} \\
&= \braket{H_{i_1} H_{i_2} \dots H_{i_{p-1}}}\braket{[H_{i_p}, H_{i_{p+1}}]}\braket{ H_{i_{p+2}} \dots H_{i_{n}}} 
\end{split}
\end{equation}
where the assumption of  statistical independence of  the matrices $H_i$ is utilized. Next, under the assumption of commutation of the first moments we get
\begin{equation}
\braket{[H_j, H_k]} = \braket{H_j} \braket{H_k} - \braket{H_k} \braket{H_j} = [\braket{H_j}, \braket{H_k}] = 0
\end{equation} 
Finally, we show that
\begin{equation}\label{lemma}
\sum_{i_1\neq \dots \neq i_n}^N \beta_{i_1,\dots,i_n}  \braket{H_{i_1} H_{i_2} \dots H_{i_n}} = 0.
\end{equation}
The number of vanishing terms in this sum, if $N > n > 1$, is equal to the number of partial permutations of length $n$ from the set of $N$ elements, i.e. $N!/(N-n)!$. Consequently, the number of all non-zero terms in the sum (\ref{termavg}) is 
\begin{equation}
G_{n-1} (N) = N^n - \frac{N!}{(N-n)!}, 
\end{equation}
where  $G_k(N)$ denotes the  $k$-th degree polynomial of the variable $N$. 
Further, we estimate that
\begin{equation}
\begin{split}
& \| \braket{H_{i_1}H_{i_2} \dots H_{i_n}} \| \le  \braket{\| H_{i_1}H_{i_2} \dots H_{i_n} \|} \\
& \le \braket{\| H_{i_1}\| \| H_{i_2} \| \dots \| H_{i_n} \|} \le K(n) / \tau^n
\end{split}
\end{equation}
and as a consequence only if $N > n > 1$ we get
\begin{equation}
\begin{split}
\| &\braket{P_n(\mathcal{H}_N)} \|  \le  \frac{\tau^n}{N^n} \| \sum_{i_1,\dots,i_n}^N \beta_{i_1,\dots,i_n} \braket{H_{i_1}H_{i_2} \dots H_{i_n}} \| \\
&\le \frac{G_{n-1} (N)}{N^n} \beta(n) K(n) = R_n(N) \beta(n) K(n). 
\end{split}
\end{equation}

\subsection{Variance} 
For the elements of the variance-matrix series $\mathrm{Var}(\tilde H_N)$ we have:
\begin{equation} \label{vars}
\begin{split}
&[S_{n,m} (\mathcal{H}_N)]_{\alpha \beta} = \\ 
&= \frac{\tau^{n+m}}{N^{n+m}} \sum_{i_1, \dots , i_{n}}^N \sum_{j_1, \dots , j_{m}}^N \beta_{i_1 \dots i_{n}} \beta_{j_1 \dots j_{m}} [S_{i_1, \dots , i_{n}, j_1, \dots , j_{m}}]_{\alpha \beta}, 
\end{split}
\end{equation}
where
\begin{equation}
\begin{split}
&[S_{i_1, \dots , i_{n}, j_1, \dots , j_{m}}]_{\alpha \beta} = \braket{\big{[}H_{i_1} \dots H_{i_{n}}\big{]}_{\alpha \beta} \big{[}H_{j_1} \dots H_{j_{m}} \big{]}_{\alpha \beta}^*} \\
&- \braket{ \big{[}H_{i_1} \dots H_{i_{n}} \big{]}_{\alpha \beta}} \braket{\big{[}H_{j_1} \dots H_{j_{m}}\big{]}_{\alpha \beta}^*}. 
\end{split}
\end{equation}
In analogy to previous considerations for the subset of indices $i_1 \neq j_1 \neq i_2 \neq j_2 \neq \dots \neq i_n \neq j_n$, under the assumption of statistical independence, we have 
\begin{equation}
\begin{split}
\sum_{i_1 \neq j_1 \neq \dots \neq i_{n} \neq j_n}^N & \beta_{i_1 \dots i_{n}} \beta_{j_1 \dots j_{m}} \times  [S_{i_1, \dots , i_{n}, j_1, \dots , j_{m}}]_{\alpha \beta} \ = 0
\end{split}
\end{equation}
for any $n+m < N$. Thus, the number of all non-zero terms in the sum (\ref{vars})  at least is 
\begin{equation}
G_{n+m-1}(N) = N^{n+m} - \frac{N!}{(N-n - m)!}.
\end{equation} 

Similarly to the earlier reasoning we estimate that
\begin{equation}
\begin{split}
    &\| S_{i_1, \dots , i_n, j_1, \dots , j_m} \| \le \braket{\| H_{i_1} H_{i_2} \dots H_{i_n} \| \| H_{j_1} H_{j_2} \dots H_{j_m} \|} \\
    &+ \braket{\|H_{i_1} H_{i_2} \dots H_{i_n} \|} \braket{\|H_{j_1} H_{j_2} \dots H_{j_m} \|} \le K(n,m) /\tau^{n+m}, 
\end{split}
\end{equation}
where we use the fact that the matrix defined by elements $[Z]_{\alpha \beta} = [X]_{\alpha \beta} [Y]_{\alpha \beta}$ satisfies the relation $\| Z \| \le \| X \| \| Y \|$. 

Finally, we obtain in analogy to prior reasoning:
\begin{equation}
\begin{split}
    \|\braket{S_{n,m}} \| &\le \frac{G_{n+m-1}(N)}{N^{n+m}} \beta(n) \beta(m) K(n,m) \\
    & = R_{n+m}(N) \beta(n)\beta(m) K(n,m). \, 
\end{split}
\end{equation}

\section{Proof of Theorem 2}
We define the  set $\mathcal{\bar H}_N = (H_N, H_{N-1}, \dots, H_1)$ which is a reverse protocol of $\mathcal{H}_N$. From the identity:
\begin{equation}
e^{-i \tilde H_{N} \tau} = e^{- \frac{i}{N} H_N \tau} \dots e^{- \frac{i}{N} H_2 \tau} e^{- \frac{i}{N} H_1 \tau} \,,
\end{equation}
we can rearrange order and get
\begin{equation}
e^{i \tilde H_{N} \tau} =  e^{\frac{i}{N} H_1 \tau} e^{\frac{i}{N} H_2 \tau} \dots e^{\frac{i}{N} H_N \tau} \,.
\end{equation}
which implies that for any polynomial $P_n(\mathcal{H}_N)$ of $n$-th degree  the  relation
\begin{equation} \label{polyrelation}
P_n(\mathcal{-H}) = - P_n(\mathcal{\bar H}_N). \,
\end{equation}
is satisfied. 
What is more, for any $n \in \mathbb{N}$, 
\begin{equation} \label{evenrelation}
\begin{split}
P_{2n}(-\mathcal{H}_N) &= P_{2n}(\mathcal{H}_N), \\
P_{2n+1}(-\mathcal{H}_N) &= -P_{2n+1}(\mathcal{H}_N)
\end{split}
\end{equation}
The joint pdf for i.i.d. matrices $\mathcal{H}_N$ has the form 
\begin{equation} \label{pdfrelation}
\begin{split}
& \rho(\mathcal{H}_N) = \varrho(H_1)\varrho(H_2) \dots \varrho(H_N) \\
&= \varrho(H_N) \dots \varrho(H_2) \varrho(H_1) = \rho(\mathcal{\bar H}_N) 
\end{split}
\end{equation}
and this leads to the relation
\begin{equation}
\braket{P_n(\mathcal{H}_N)} = \braket{P_n(\mathcal{\bar H}_N)}.
\end{equation}

Finally, taking into consideration Eqs. (\ref{polyrelation}), (\ref{evenrelation}) and (\ref{pdfrelation}), we have 
\begin{equation}
\braket{P_{2n} (\mathcal{H}_N)} = - \braket{P_{2n} (\mathcal{\bar H}_N)} = - \braket{P_{2n} (\mathcal{H}_N)} = 0
\end{equation}
for any $n \in \mathbb{N}$. 
Similarly to before, we have the relation
\begin{equation}
S_{n,m}(-\mathcal{H}_N) = S_{n,m}({\mathcal{\bar H}_N}).
\end{equation}
Thus, for any $n,m \in \mathbb{N}$ with pdf Eq. (\ref{pdfrelation}), in analogy one can shown that
\begin{equation}
\begin{split}
    S_{2n,2m+1} (\mathcal{H}_N) = S_{2n+1,2m} (\mathcal{H}_N) = 0 
\end{split}
\end{equation}
and this proves the first part of Theorem \ref{theorem1}. 

The proof of  the second part is straightforward if one notices that for even pdf $\varrho(H)$ we have
\begin{equation}
\braket{P_n(\mathcal{-H})} = \braket{P_n(\mathcal{H}_N)}. \,
\end{equation}

\section{Variance of unitary time-evolution}

Let us assume that $x_1, x_2, \dots$ are complex random variables  where each of them behaves as 
\begin{eqnarray}
 \mathrm{Var}(x_k) = O(\frac{1}{N}). \,
\end{eqnarray}
Then the sum of them 
\begin{eqnarray} 
 \mathrm{Var}(\sum_i x_i) = \sum_{i,j} \mathrm{Cov} (x_i, x_j) \le \sum_{i,j} \sqrt{\mathrm{Var}(x_i) \mathrm{Var}(x_j)} \nonumber \\
\end{eqnarray}
behaves as  
\begin{eqnarray}\label{randsum}
 \mathrm{Var}(\sum_i x_i) = O(\frac{1}{N}). \,
\end{eqnarray}
Next, we would like to estimate  the variance of the product. To this aim,  let us define the centered random variable:
\begin{eqnarray}
\delta x_i = x_i - \braket{x_i}
\end{eqnarray}
where $\mathrm{Var}(\delta x_i) = \mathrm{Var}(x_i)$ and $\braket{\delta x_i} = 0$. Then the product can be expanded as
\begin{eqnarray} 
 \prod_i x_i = \prod_i \braket{x_i} + \sum_k \delta x_k \prod_{i\neq k} \braket{x_i} + \dots
\end{eqnarray}
Thus up to the leading orders of $N$ we have
\begin{eqnarray} 
 \mathrm{Var}(\prod_i x_i) &=&  \sum_{k,m} \abs{\braket{x_k}}\abs{\braket{x_m}} \prod_{i \neq k, m} \abs{\braket{x_i}}^2 \mathrm{Cov} (x_k, x_m)     \nonumber \\
 &+& \dots
\end{eqnarray}
which implies
\begin{eqnarray} \label{randprod}
 \mathrm{Var}(\prod_i x_i) = O(\frac{1}{N}). \,
\end{eqnarray}

For the self-averaging effective Hamiltonian we shown that any of its elements is asymptotically equivalent to functions belonging to $O(1/N)$ and since the unitary matrix could  be expressed as a series
\begin{eqnarray}
 U_N(\tau,0) = \sum_{k=1} \frac{(- i \tilde H_N \tau)^k}{k!}
\end{eqnarray}
which involves sums and products of the effective Hamiltonian elements, hence we conclude also that:
\begin{eqnarray} \label{univar}
 \| \mathrm{Var}[U_N (\tau,0)] \| = O(\frac{1}{N}). \,
\end{eqnarray}

\section{Mean of unitary time-evolution}

We want to prove that 
\begin{equation} \label{toprove}
\|\braket{T e^{-i \int_0^\tau dt H_N(t)}} - T e^{-i \int_0^\tau dt \braket{H_N(t)}}] \| = O(\frac{1}{N})
\end{equation}
assuming that 
\begin{equation}
\|\braket{\tilde H_N} - \widehat H_N \| = O(\frac{1}{N}), 
\end{equation}
where $\widehat H_N = \frac{1}{\tau} \int_0^\tau dt \braket{H_N(t)}$
and
\begin{eqnarray}
[\braket{H(t)}, \braket{H(s)}] = 0
\end{eqnarray}
for any $t,s \in \mathcal{T}$. First, one can estimate:
\begin{eqnarray} \label{est}
&&\|\braket{T e^{-i \int_0^\tau dt H_N(t)}} - T e^{-i \int_0^\tau dt  \braket{H_N(t)}} \| = \nonumber \\
&=& \|\braket{e^{-i \tilde H_N \tau} - e^{-i \widehat H_N \tau }} \| \le \sum_k \frac{\tau^k}{k!} \|\braket{ \tilde H_N^k - \widehat H_N^k}] \| \nonumber \\
\end{eqnarray}
Let us define the  matrix
\begin{eqnarray}
\delta H_N = \tilde H_N - \widehat H_N. 
\end{eqnarray}
We note that it satisfies 
\begin{eqnarray} \label{deltamean}
\|\braket{\delta H_N}\| = \| \braket{\tilde H_N - \widehat H_N} \| = O(\frac{1}{N}), \,
\end{eqnarray}
and variance 
\begin{eqnarray} \label{vardelta}
\|\mathrm{Var}(\delta H_N)\| = \|\mathrm{Var}(\tilde H_N )\| = O(\frac{1}{N}). \,
\end{eqnarray}
Further, 
\begin{eqnarray}
\tilde H_N^k - \widehat H_N^k = \sum_{m=0}^{k-1} \widehat H_N^m \delta H_N \widehat H_N^{k-m-1} + \dots
\end{eqnarray}
where dropped terms involve higher powers of $\delta H_N$ elements. According to (\ref{deltamean}) and (\ref{vardelta}), the leading order of the average is then equal to:
\begin{eqnarray}
\braket{\tilde H_N^k - \widehat H_N^k} = \sum_{m=0}^{k-1} \widehat H_N^m \braket{\delta H_N} \widehat H_N^{k-m-1} + \dots
\end{eqnarray}
However, due to the sub-multiplicative condition of the norm we conclude that
\begin{eqnarray}
\| \sum_{m=0}^{k-1} \widehat H_N^m \braket{\delta H_N} \widehat H_N^{k-m-1} \| \le k \| \braket{\delta H_N} \| \| \widehat H_N\| ^{k-1} \nonumber \\
\end{eqnarray}
and this finally implies 
\begin{eqnarray}
\| \braket{\tilde H_N^k - \widehat H_N^k} \| = O(\frac{1}{N}). 
\end{eqnarray}
According to (\ref{est}), this proves the relation (\ref{toprove}). 

\section{Beyond the convergence domain}

\subsection{Variance $S_N$}
Let us consider the MRQ evolution
\begin{eqnarray}
    U_N(\tau, 0) = T e^{-i \int_0^\tau dt H_N(t)}  = \prod_{k=1}^N e^{-\frac{i}{N} H_k \tau},   
\end{eqnarray}
where the convergence condition (\ref{convcond}) is not satisfied. Nevertheless, one can always split the unitary evolution into $m$ products
\begin{eqnarray}
    U_N(\tau, 0) = \prod_{k=1}^m U_{N/m} (\tau_{k+1}, \tau_k), 
\end{eqnarray}
where $\tau_k = (k-1) \tau / m$, such that for each term the convergence condition is obeyed. 
Then, if 
\begin{eqnarray}
\mathrm{Var}[U_{N/m} (\tau_{k+1}, \tau_k)] = O(\frac{1}{N}) 
\end{eqnarray}
according to relations (\ref{randsum}) and (\ref{randprod}), one concludes that also an arbitrary finite product of matrices satisfies
\begin{eqnarray}
\| \mathrm{Var}[U_N(\tau, 0)] \| = \| \mathrm{Var}[\prod_{k=1}^m U_{N/m} (\tau_{k+1}, \tau_k)] \| = O(\frac{1}{N}). \nonumber \\
\end{eqnarray}
\subsection{Distance $D_N$}
Further let us define a matrix
\begin{eqnarray}
    \delta U_{N/m}^{(k)} = U_{N/m} (\tau_{k+1}, \tau_k) - \widehat U_{N/m} (\tau_{k+1}, \tau_k), 
\end{eqnarray}
where 
\begin{eqnarray} 
    \widehat U_{N/m} (\tau_{k+1}, \tau_k) = T e^{-i \int_{\tau_{k+1}}^{\tau_k} dt \braket{H_N(t)}} .
\end{eqnarray}
If $H(t)$ commute in the statistical sense then from Eq. (\ref{dn}) we obtain
\begin{eqnarray} \label{uvar}
\|\braket{\delta U_N^{(k)}}\| = O(\frac{1}{N}) \,
\end{eqnarray}
and variance 
\begin{eqnarray}
\|\mathrm{Var}(\delta U_N^{(k)})\| = \|\mathrm{Var} [U_{N/m} (\tau_{k+1}, \tau_k)]\| =  O(\frac{1}{N}). \,
\end{eqnarray}
Finally, we can expand the unitary time-evolution matrix up to leading terms of $N$:
\begin{eqnarray}
    && U_N(\tau, 0) = \widehat U_{N} (\tau, 0) \nonumber \\
    &+&  \sum_{k=0}^{m-1} \widehat U_{kN/m} (\tau, \tau_{m-k+1})  \delta U_{N/m}^{(k)} \widehat U_{(m-k-1)N/m} (\tau_{m-k}, 0) \nonumber \\
    &+& \dots
\end{eqnarray}
which due to (\ref{uvar}) once again gives
\begin{equation}
    \| \braket{U_N(\tau, 0)} - \widehat U_{N} (\tau, 0) \| = O(\frac{1}{N})
\end{equation}
and proves Eq. (\ref{dn}).




\section*{ACKNOWLEDGMENTS}
The work supported by  the Grant No. NCN 2015/19/B/ST2/02856.

\bibliographystyle{plainnat}

\end{document}